\documentclass[12pt,preprint]{aastex}
\usepackage{longtable}
\def\Deg{\hbox{${}^\circ$\llap{.}}}

\def\Sec{\hbox{${}^{\prime\prime}$\llap{.}}}
\def\kms{$\mathrm{km\;s}^{-1}$} 
\def\dg{^\circ} 
\def\ha{H$\alpha$}

\def\nii{[N~{\small II}]} 
\def\niip{[N~{\small II}]$\,\lambda6548$} 
\def\niig{[N~{\small II}]$\,\lambda6583$} 
\def\niipg{[N~{\small II}]$\,\lambda\lambda6548,6583$}
\def\sii{[S~{\small II}]} 
\def\siip{[S~{\small II}]$\,\lambda6716$} 
\def\siig{[S~{\small II}]$\,\lambda6731$} 
\def\siipg{[S~{\small II}]$\,\lambda\lambda6716,6731$} 

\def\msun{M$_{\odot}$} 
\def\mbh{$M_{\bullet}$} 
\def\sigmac{$\sigma_c$}

\def\mlstar{$(M/L)_\star$}
\def\mlsun{(M/L)$_\odot$}
\def\lbulge{$L_{\it bulge}$}
\def\hst{{\it HST\/}}

\def\h3{$h_{3}$}
\def\h4{$h_{4}$}

\shorttitle{SBHs in Brightest Cluster Galaxies}
\shortauthors{Elena Dalla Bont\`a et al.}
\begin{document}

\title{The High-Mass End of the Black Hole Mass Function: Mass
  Estimates in Brightest Cluster Galaxies\footnote{Based on
  observations made with ESO telescopes at the La Silla Paranal
  Observatory under programme ID 279.B-5004(A).}}

\author{E. Dalla Bont\`a}
\affil{Dipartimento di Astronomia, Universit\`a degli Studi di Padova, Padova, Italy}
\affil{Herzberg Institute of Astrophysics, Victoria, CA}

\email{elena.dallabonta@unipd.it}
\author{L. Ferrarese}
\affil{Herzberg Institute of Astrophysics, Victoria, CA}
\email{laura.ferrarese@nrc-cnrc.gc.ca}

\author{E.~M. Corsini}
\affil{Dipartimento di Astronomia, Universit\`a degli Studi di Padova, Padova, Italy}
\email{enricomaria.corsini@unipd.it}

\author{J. Miralda-Escud\'e}
\affil{Institut de Ci\`encies de l'Espai (IEEC-CSIC/ICREA)}
\email{miralda@aliga.ieec.uab.es}

\author{L. Coccato}
\affil{Max-Planck-Institut fuer extraterrestrische Physik, Garching bei Muenchen, Germany}
\email{lcoccato@mpe.mpg.de}

\author{M. Sarzi}
\affil{Centre for Astrophysics Research, University of Hertfordshire,
  Hatfield, UK}
\email{m.sarzi@herts.ac.uk}

\author{A. Pizzella}
\affil{Dipartimento di Astronomia, Universit\`a degli Studi di Padova, Padova, Italy}
\email{alessandro.pizzella@unipd.it}

\and

\author{A. Beifiori}
\affil{Dipartimento di Astronomia, Universit\`a degli Studi di Padova, Padova, Italy}
\email{alessandra.beifiori@unipd.it}

\begin{abstract}

We present Hubble Space Telescope imaging and spectroscopic
observations of three Brightest Cluster Galaxies, Abell 1836-BCG,
Abell 2052-BCG, and Abell~3565-BCG, obtained with the Wide Field and
Planetary Camera 2, the Advanced Camera for Surveys and the Space
Telescope Imaging Spectrograph. The data provide detailed information
on the structure and mass profile of the stellar component, the dust
optical depth, and the spatial distribution and kinematics of the
ionized gas within the innermost region of each galaxy. Dynamical
models, which account for the observed stellar mass profile and
include the contribution of a central supermassive black hole (SBH),
are constructed to reproduce the kinematics derived from the H$\alpha$
and [N II]$\lambda\lambda$6548,6583 emission lines. Secure SBH
detection with $M_{\bullet} = 3.61^{+0.41}_{-0.50} \times 10^9$ M$_{\odot}$
and $M_{\bullet} = 1.34^{+0.21}_{-0.19} \times 10^9$ M$_\odot$,
respectively, are obtained for Abell 1836-BCG and Abell 3565-BCG,
which show regular rotation curves and strong central velocity
gradients. In the case of Abell~2052-BCG, the lack of an orderly
rotational motion prevents a secure determination, although an upper limit of
$M_{\bullet} \lesssim 4.60 \times 10^9$ M$_{\odot}$ can be placed on the 
mass of the central SBH. These
measurements represent an important step forward in the
characterization of the high-mass end of the SBH mass function.

\end{abstract}

\keywords{black hole physics --- galaxies: elliptical and lenticular,
cD --- galaxies: nuclei --- galaxies: kinematics and dynamics}

\section{Introduction}\label{sec:intro}

Within the past decade, the focus in supermassive black holes (SBHs)
studies has moved from the dynamical measurement of SBH masses,
$M_{\bullet}$, in nearby galaxies (see review by Ferrarese \& Ford
2005), to the characterization of scaling relations connecting
$M_{\bullet}$ to the large scale properties of their hosts (Kormendy
\& Richstone 1995; Ferrarese \& Merritt 2000; Gebhardt et al. 2000;
Graham et al. 2001; Ferrarese 2002; Marconi \& Hunt 2003). Such relations, combined
with the knowledge of the galaxy luminosity or velocity dispersion
functions, lead to a direct determination of the local SBH mass
function and, by comparison with the energetics of high redshift AGNs,
accretion history (e.g. Marconi et al. 2004; Shankar et al. 2004;
Benson et al. 2007; Tundo et al. 2007; Graham et al. 2007; Lauer et
al. 2007). Furthermore, the tightness of the relations linking
galactic properties to $M_{\bullet}$ is indicative of a
formation/evolutionary history in which SBHs and galaxies are causally
connected. Indeed, feedback from AGN activity is believed to play an
important, perhaps even fundamental role in the evolution of galaxies
(e.g. Binney \& Tabor 1995; Suchkov et al. 1996; Ciotti \& Ostriker
2001; Schawinski et al.2006; Springel et al. 2005; Hopkins et
al. 2007; Di Matteo et al. 2007).

In this context, Brightest Cluster Galaxies (BCGs) are of particular
interest. Their privileged location at the center of a massive cluster
implies that they have undergone a particularly extensive merging
history (Khochfar \& Silk 2006) and are likely to host the most
massive SBHs in the local Universe (Yoo et al. 2007). The latter
conclusion is also supported by all scaling relations which are known
to be obeyed by local SBHs, that predict the most massive SBHs to
reside in most massive galaxies. As such, BCGs constitute an excellent
laboratory to search for the local relics of the most powerful
high-redshift quasars (Willott et al. 2003; Vestergaard 2004), and to
investigate the role of mergers in the black hole mass function.

Unfortunately, direct dynamical measurements of SBHs masses in BCGs
are exceedingly difficult - with only two such measurements made
to-date (in M87 and NGC 1399, Harms et al. 1994; Macchetto et
al. 1997; Houghton et al. 2006). The reason for this is simple: SBH
mass measurements based on resolved kinematic tracers (gas or stars)
need to be carried out within the SBH ``sphere of influence", i.e. the
region of space within which the SBH dominates the overall
gravitational potential. For a $10^9$ $M_{\odot}$ SBH, the sphere of
influence becomes unresolved for optical measurements with the Hubble
Space Telescope at distances beyond $\sim 100$ Mpc (see Figure 44 of
Ferrarese \& Ford 2005). Few BCGs are located within this limit.  To
compound the problem, bright ellipticals are characterized by shallow
stellar density profiles and faint central surface brightnesses (e.g.,
Ferrarese et al. 1994; Lauer et al. 1995; Rest et al. 2001; Ferrarese
et al. 2006) making the detection of stellar absorption lines with
{\it HST} prohibitively expensive.  The latter difficulty can be
overcome with the use of gas dynamics. Emission lines (mainly \ha\ and
\nii) from the gas are bright and easily measured. If the gas is
confined in a disk, there is little ambiguity in the velocity
distribution (Ho et al. 2002), and since the method was first used
(Harms et al. 1994; Ferrarese et al. 1996; Ferrarese \& Ford 1999) an
increasing amount of attention has been devoted to the theoretical
aspects of the dynamical modeling (Maciejewski \& Binney 2001; Barth
et al.  2001; Cappellari et al. 2002; Coccato et al. 2006).

The lack of a secure characterization of the SBH mass function above
the $10^9$ $M_{\odot}$ mark is troublesome.  Lauer et al. (2007)
suggest that the relations between $M_{\bullet}$ and host bulge
luminosity, $L_{bulge}$, (Kormendy \& Richstone 1995) and velocity
dispersion (Ferrarese \& Merritt 2000; Gebhardt et al. 2000) would
predict significantly different $M_{\bullet}$ if extrapolated above
$10^9$ $M_{\odot}$. In particular, the $M_{\bullet} - \sigma$ relation
would predict less massive SBHs in BCGs than the $M_{\bullet} -
L_{bulge}$ relation, due to the slower increase of $\sigma$ with
galaxy luminosity observed for BCGs compared to the bulk of the
Early-Type galaxy population (Bernardi et al. 2007). The difference in
the predicted mass is such to significantly affect (by an order of
magnitude) the high-mass end of the local SBH mass function.  von der
Linden et al. (2007) argue that the shallower dependence of $\sigma$
on $L$ applies to BCG but not to comparably massive non-BCG galaxies,
which instead follow the canonical Faber-Jackson relation defined by
less massive systems. This implies that BCGs and non-BCG galaxies of
comparable mass must occupy a different locus in either, or both, the
$M_{\bullet} - \sigma$ and $M_{\bullet} - L_{bulge}$. This result is
in contrast with the findings of Batcheldor et al. (2007), who argue
that SBHs masses predicted from NIR luminosities (Marconi \& Hunt
2003) are consistent with those predicted from $\sigma$. They
attributed the discrepancy noted by Lauer et al. (2007) (who used
$V-$band luminosities) to a bias introduced by the inclusion, when
computing $L_{bulge}$, of extended blue envelopes around BCGs. At
present these ambiguities prevent us from testing any theoretical
prediction on the high-mass end of the black hole mass function based on
the observed quasar abundances and merger histories.

In this paper, we analyze the kinematics of the ionized gas in the
nuclear region of three BCGs -- Abell~1836-BCG, Abell~2052-BCG, and
Abell~3565-BCG -- in order to constrain the mass of the central
SBHs. The data were obtained using the Space Telescope Imaging
Spectrograph (STIS) on board the Hubble Space Telescope (HST),
supplemented with imaging data from the Advanced Camera for Surveys
(ACS) and the Wide Field and Planetary Camera 2 (WFPC2).  Furthermore,
the central stellar velocity dispersion of Abell 1836-BCG was obtained
from ground based spectroscopy.  The paper is organized as
follows. The criteria of galaxy selection are presented in \S
\ref{sec:sample}. ACS and STIS observations are described and analyzed
in \S\S \ref{sec:imaging} and \ref{sec:spectroscopy}, respectively.
Ground based spectroscopic observations of Abell 1836-BCG are
described and analyzed in \S \ref{sec:groundspectrospcopy}. 

The SBH masses of Abell~1836-BCG and Abell~3565-BCG, and an upper
limit for the SBH mass of Abell~2052-BCG are derived in \S
\ref{sec:model}. In \S \ref{sec:conclusions} results are compared to
the predictions of the SBHs scaling laws and conclusions are given.

\section{Galaxy sample}
\label{sec:sample}

Naturally, the place to search for the most massive black holes in the
universe is in the most nearby and most massive BCGs, where the
problems of angular resolution to measure black hole masses are
minimized.  The galaxies discussed in this contribution were selected
from the BCG sample observed by Laine et al. (2003) with the HST Wide
Field and Planetary Camera 2 (WFPC2). Dynamical studies in giant
ellipticals are performed more effectively by modeling gas, rather
than stellar kinematics (see discussion in Ferrarese \& Ford 2005),
both because of the low stellar surface brightness that characterizes
the core regions of such systems (e.g. Ferrarese et al. 2006) and
because when applied to slowly-rotating, pressure supported galaxies,
stellar dynamical models depend heavily on the knowledge of the
anisotropy in the stellar velocity dispersion, a difficult quantity to
constrain observationally (e.g. van der Marel 1998).  Eleven of the 81
galaxies in the Laine et al. (2003) sample appear to contain well
defined nuclear dust structures, which can trace regular gas
kinematics (Ho et al. 2002). After selecting the most regular nuclear
dust structures among those eleven objects and after excluding
galaxies for which existing ground based spectra showed no evidence of
nuclear \ha\ and \niipg\ in emission, we selected the four BCGs
predicted to host the most massive SBHs (based on the correlations
between black hole mass, $M_{\bullet}$, and the host bulge stellar
velocity dispersion, $\sigma$, and/or $B$-band luminosity, $L_{B}$)
and for which the SBH sphere of influence could be well resolved using
the 0\Sec1 wide slit of HST's Space Telescope Imaging Spectrograph
(STIS).
 
One galaxy (Abell~2593-BCG) was not observed due to the premature
failure of STIS. The sample discussed in this paper therefore includes
three BCGs, namely Abell~1836-BCG (PGC~49940), Abell~2052-BCG
(UGC~9799) and Abell~3565-BCG (IC~4296). The main properties of these
galaxies are summarized in Table~\ref{tab:sample}.
\section{Observations, data reduction, and analysis: Imaging}
\label{sec:imaging}

\subsection{Observations and data reduction}
\label{sec:acs}

The luminosity density and ionized gas distribution of the target
galaxies, both of which are required for the dynamical modeling
described in \S \ref{sec:model}, have been constrained using HST ACS and
WFPC2 images. Basic information about the data can be found in
Table~\ref{tab:ACSlog}.

The ACS images were obtained with the High Resolution Channel (HRC) as
part of program GO-9838 (P.I. L. Ferrarese). The HRC detector is a
$1024\times1024$ SITe CCD with a plate scale of
$0\Sec028\times0\Sec025$ pixel$^{-1}$. Because of the large, but well
characterized, geometric distortion affecting the ACS, the
$29''\times26''$ HRC field of view projects on the plane of the sky as
a rhomboid with x and y axes forming a 84\Deg2 angle.  Each galaxy was
observed with two filters: F435W (which approximates the Johnson $B$-band) and the narrow-band ramp filter FR656N, tuned to
cover a 130 \AA~ wide spectral region centered on the redshifted \ha\
and \niipg\ emission expected at the nuclear location.  To help in
identifying and correcting cosmic ray events, two back-to-back
exposures were taken with each filter.

The WFPC2 images, all employing the F814W filter (similar to the
Kron-Cousin $I$ band), belong to programs GO-5910 (P.I. T. Lauer) and
GO-8683 (P.I. R.P. van der Marel) and were downloaded from the public
HST archive.  Two images with a total exposure time of 1000 s are
available for each Abell~1836-BCG and Abell~2052-BCG, while ten images
are available for Abell~3565-BCG, for a total 11,600 s exposure. In
all cases, the galaxies are centered on the higher resolution
Planetary Camera (PC), which consists of an $800\times800$ pixels
Loral CCD with a plate scale of $0\Sec0455$ pixel$^{-1}$, yielding a
nominal field of view of $36'' \times36''$.  For both ACS and WFPC2
images, the telescope was guiding in fine lock, giving a typical rms
tracking error per exposure of 0\Sec005.

All images were calibrated using the standard reduction pipelines
maintained by the Space Telescope Science Institute (PyRAF/CalACS for
the ACS images and IRAF/CalWFPC for the WFPC2 images). Reduction steps
include bias subtraction, dark current subtraction and flat-fielding,
as described in detail in the ACS and WFPC2 instrument and data
handbooks (Baggett et al. 2002; Heyer et al.2004; Pavlovsky et
al. 2004a,b).

Subsequent analysis was performed using standard tasks in {\tt
IRAF}\footnote{IRAF is distributed by the National Optical Astronomy
Observatories which are operated by the Association of Universities
for Research in Astronomy (AURA) under cooperative agreement with the
National Science Foundation.}.  For each galaxy, the alignment of
images obtained with the same instrument and filter was checked by
comparing the centroids of stars in the field of view; in all cases
but one, the alignment was found to be better than a few hundredths of
a pixel and the images were combined without further processing using
{\tt IRAF/IMCOMBINE}. The exception are the WFPC2 images of Abell
3565-BCG, which were dithered by $\approx1''$ relative to one
another. These images were aligned using {\tt IMSHIFT} prior to being
combined.

In combining the images, pixels deviating by more than three times the
local sigma -- calculated from the combined effect of Poisson and
read-out noise -- were flagged as cosmic rays and rejected.  Finally,
the resulting ACS images were corrected for geometric distortion using
{\tt PYRAF/PYDRIZZLE}.

\begin{figure}
\epsscale{1.}
\plotone{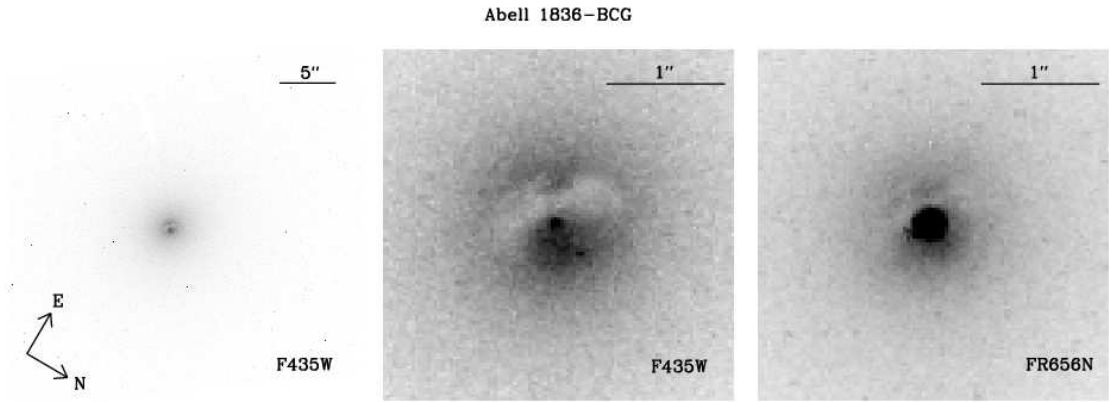}
\caption{{\small ACS/HRC images of Abell~1836-BCG.  The {\em panels}
show the F435W full frame ({\em left}) and a zoom toward the center
({\em center}).  The same central section is shown in the {\em right
panel} for the FR656N image.  The grey-scale used for the left panel
(full frame) is kept the same in Figures~\ref{acs1836}-\ref{acs3565},
while the gray-scale for the other panels changes from filter to
filter in order to highlight the distribution of the nuclear dust. All
images have been background subtracted.}\label{acs1836}}
\end{figure}

\begin{figure}
\epsscale{1.}
\plotone{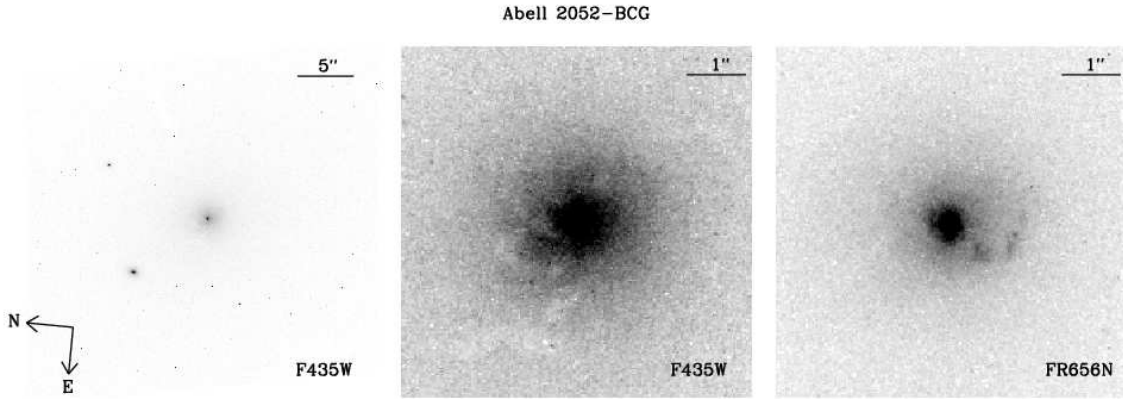}
\caption{{\small ACS/HRC images of  Abell~2052-BCG. See
    Figure~\ref{acs1836} for details.}
\label{acs2052}}
\end{figure}

\begin{figure}
\epsscale{1.}
\plotone{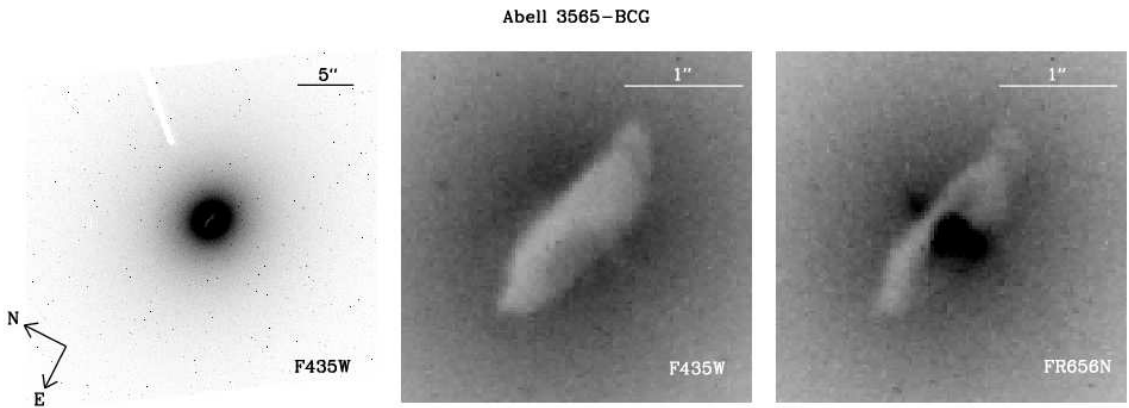}
\caption{{\small ACS/HRC images of  Abell~3565-BCG. See
    Figure~\ref{acs1836} for details.}
\label{acs3565}}
\end{figure}

\begin{figure}
\epsscale{1.}
\plotone{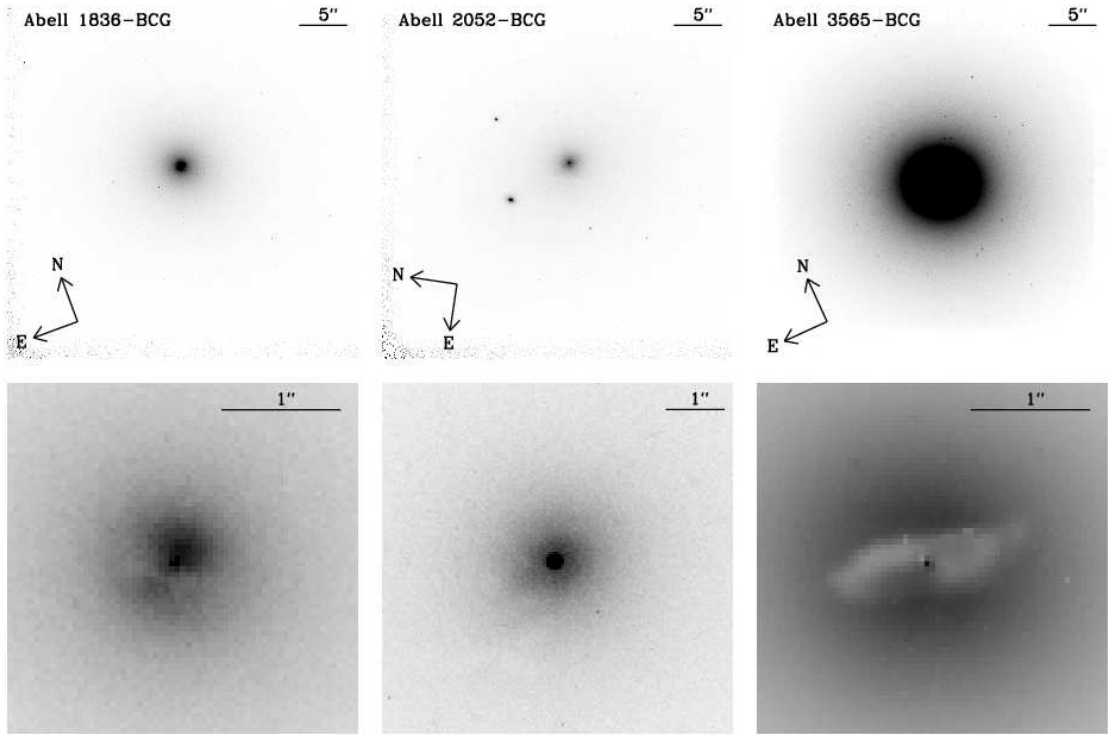}
\caption{{\small WFPC2/F814W images of Abell~1836-BCG ({\em left
    panels}), Abell~2052-BCG ({\em middle panels}), and Abell~3565-BCG
    ({\em right panels}).  The {\em top panels} show the full frame,
    while the {\em bottom panels} show a zoom toward the center. The
    grey-scale used for the top (full frame) images is kept the same for all
    the three galaxies, while the gray-scale for the central sections
    changes from galaxy to galaxy to highlight the distribution of the
    nuclear dust. All images have been background
    subtracted.}\label{wfpc2}}
\end{figure}

The final, calibrated ACS and WFPC2 images of each galaxy are shown in 
Figures~\ref{acs1836}-\ref{acs3565} and \ref{wfpc2},
respectively.

\subsection{Isophotal analysis}
\label{sec:phot}

Isophotal parameters (coordinates $X,Y$ of the isophotal center;
surface brightness profile $I(\phi)$; ellipticity $\epsilon$; major
axis position angle $\theta$; and deviations of the isophotes from
pure ellipses) were measured using the {\tt IRAF} task {\tt ELLIPSE}
(Jedrzejewski 1987). For each semi-major axis length, the best fitting
set of parameters $X,~Y,~\epsilon$ and $\theta$ are those that
minimize the sum of the squares of the residuals between the data and
the first two moments of the Fourier series expansion of the
azimuthally-sampled intensity

\begin{equation}\label{fourier}
I(\phi)=I_0+\sum_k[A_k sin(k\phi)+B_k cos(k\phi)]
\end{equation}

The third and fourth order moments ($A_3,~B_3,~A_4$ and $B_4$)
describe three and four-fold deviations of the isophotes from pure
ellipses, respectively.  The $B_4$ term in particular describes the
shape by distinguishing boxy ($B_4 < 0$) from disky ($B_4
> 0$) galaxies.  

\begin{figure}
\epsscale{1.}
\plotone{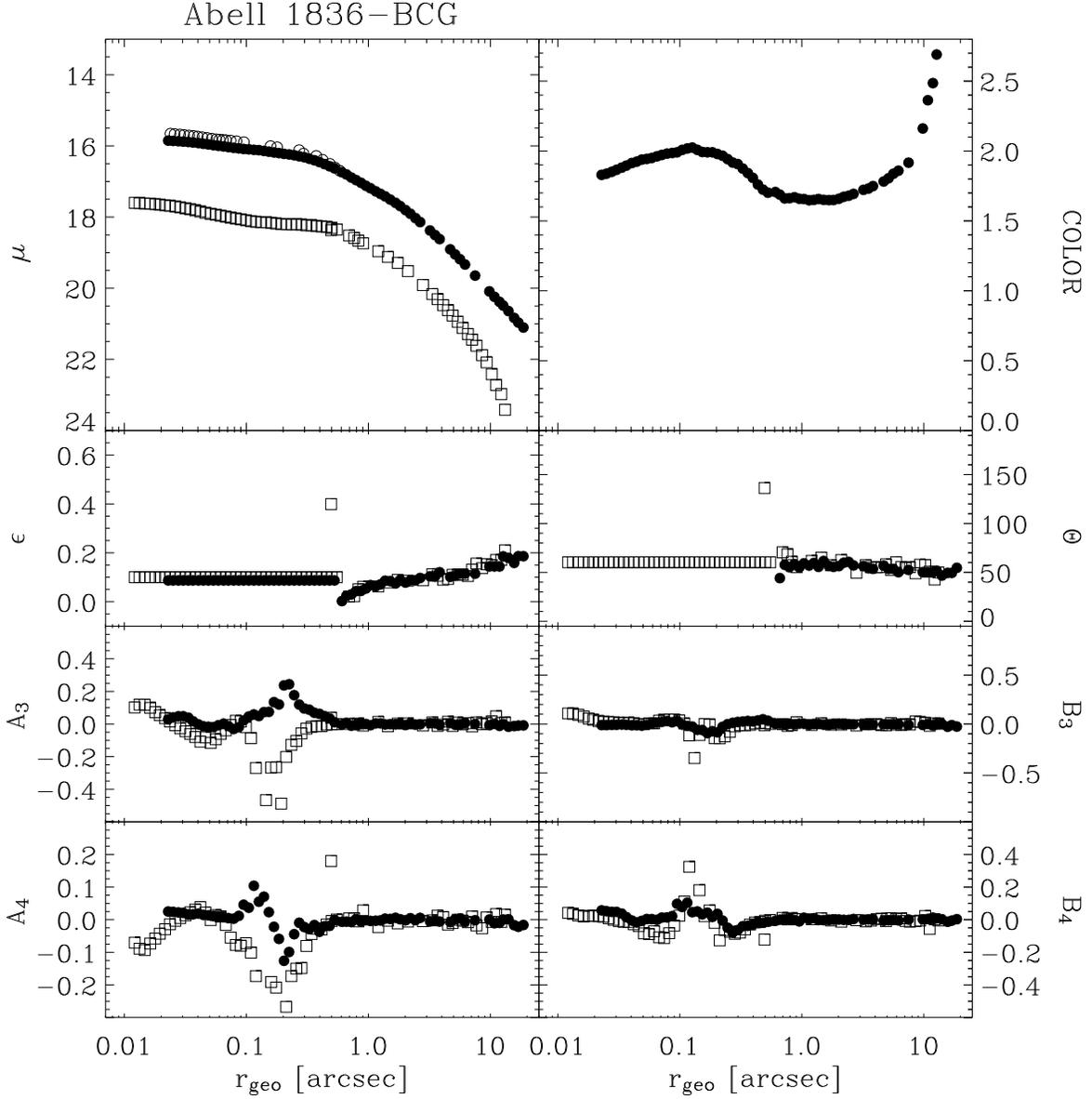}
\caption{{\small Isophotal parameters for Abell 1836-BCG plotted
    against the ``geometric mean'' radius
    $r_{geo}=a(1-\epsilon)^{1/2}$, with $a$ measured along the
    semimajor axis of the galaxy. The panels show the radial profiles
    of: surface brightness $\mu$ (in mag arcsec$^{-2}$) in ACS/F435W
    (squares) and WFPC2/F814W (filled circles) bands, F435W$-$F814W
    (filled circles) color, ellipticity $\epsilon$, position angle
    $\theta$ (in degrees, measured from north to east), and parameters
    $A_3$, $A_4$, $B_3$, and $B_4$, measuring deviations of the
    isophotes from pure ellipses (see text for details).  Open circles
    show the inner radial profile of $\mu$ measured on the WFPC2/F814W
    image corrected for dust obscuration (as described in the text)
    with the addition of the Galactic extinction correction by
    Schlegel et al. (1998).}
\label{phot1836}}
\end{figure}

\begin{figure}
\epsscale{1.}
\plotone{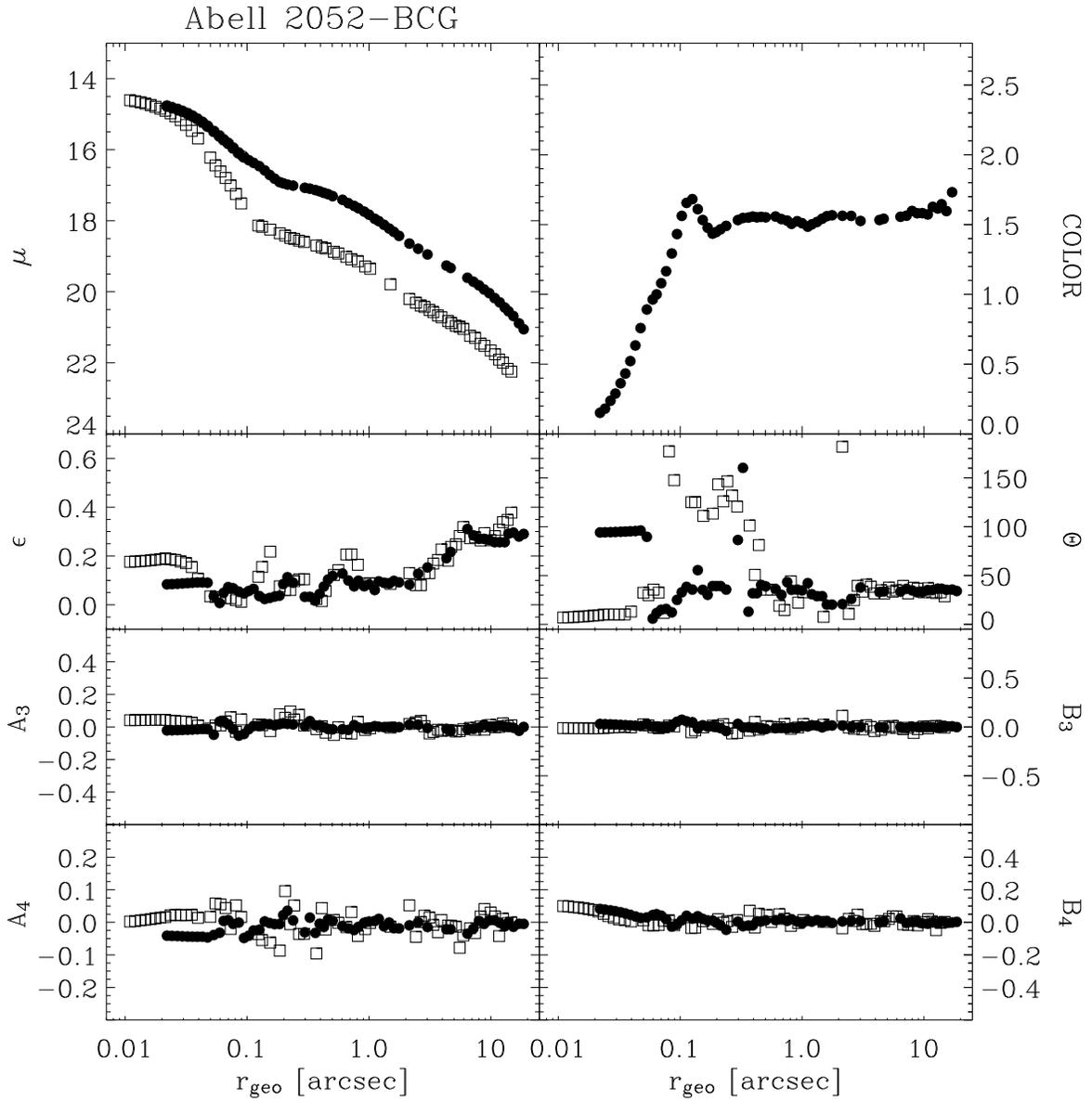}
\caption{{\small The same as for Figure~\ref{phot1836}, but for Abell 2052-BCG.}
\label{phot2052}}
\end{figure}

\begin{figure}
\epsscale{1.}
\plotone{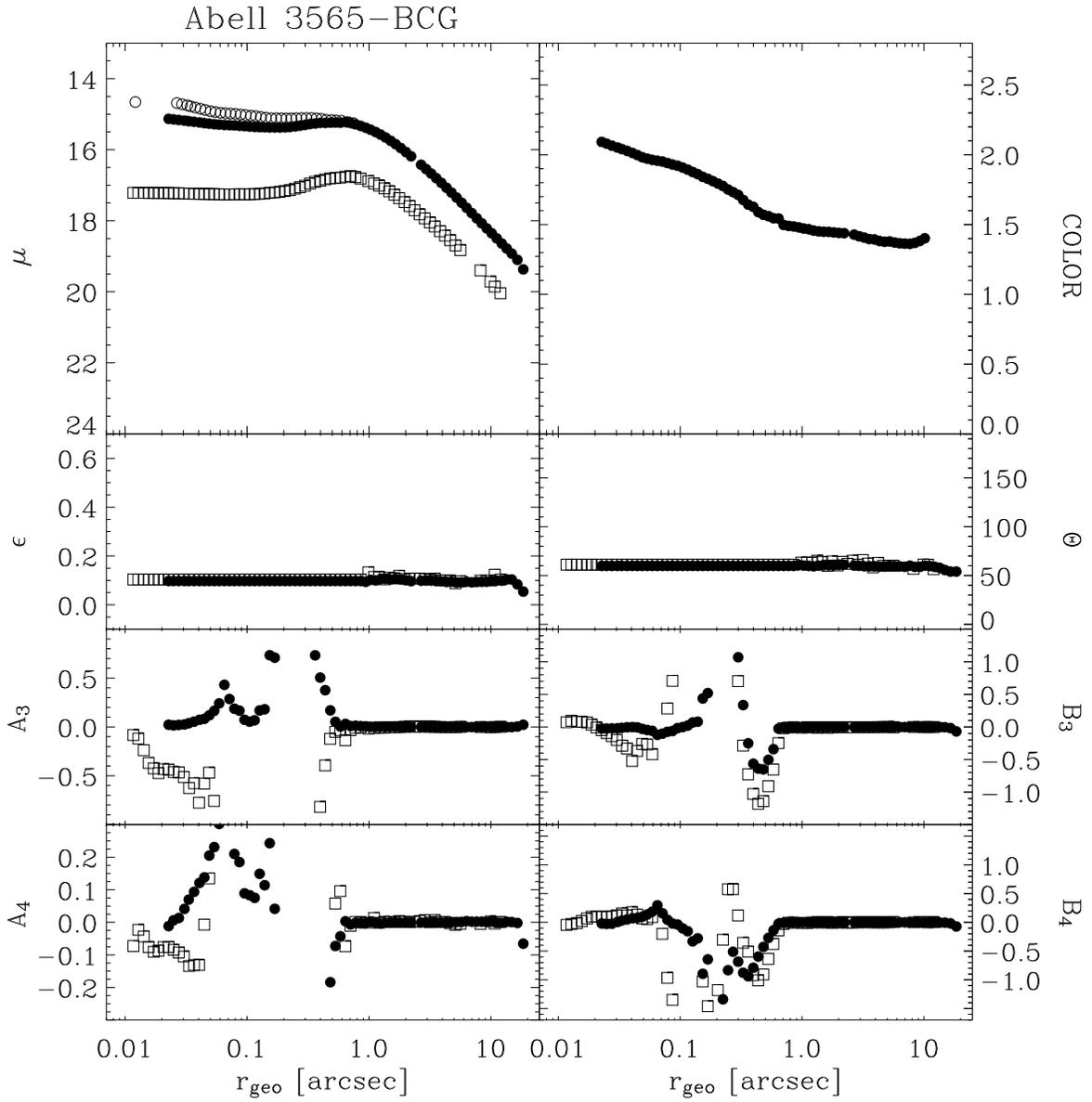}
\caption{{\small The same as for Figure~\ref{phot1836}, but for Abell 3565-BCG.}
\label{phot3565}}
\end{figure}

The isophotal semi-major axis was increased logarithmically, with each
isophote being fitted at a semi-major axis 10\% longer than that of
the isophote preceding it.  The dust structure in Abell~2052-BCG is
localized enough not to affect significantly the recovery of the
underlying stellar distribution, so all parameters were allowed to
vary freely while fitting the isophotes.  For the other galaxies, a
fully unconstrained isophotal solution could be found only in the
regions unaffected by dust ($r >$ 0\Sec6 and $r> $ 0\Sec9 for Abell
1836-BCG and Abell 3565-BCG, respectively). In the areas affected by
dust, the isophotes were fitted by fixing both ellipticity and
position angle to the average values obtained in the dust-unaffected
region ( 0\Sec6 $< r < 10''$ and 0\Sec9 $< r < 10''$ for Abell
1836-BCG and Abell 3565-BCG, respectively). In all cases, pixels
deviating by more than three times the standard deviation of the mean
intensity along each trial isophote were excluded from the fit,
thereby avoiding contamination by foreground stars, companion galaxies
(as in the case of Abell 2052-BCG), globular clusters, and bad
pixels. All isophotal parameters are plotted in Figures
~\ref{phot1836}--\ref{phot3565} as a function of the ``geometric
mean'' radius $r_{geo}$, defined as $a[1-\epsilon(a)]^{1/2}$, $a$
being the semi-major axis length.

Conversion of $I(\phi)$ (which is given in counts per seconds) to
surface brightness in the AB magnitudes system was performed following
Holtzman et al. (1995) for the WFPC2 images, and Sirianni et
al. (2005) for the ACS images:

\begin{equation}
{\mu}_{F814W}= -2.5\,\log I_s(\phi)_{\rm F814W}+14.998,
\end{equation}
\begin{equation}
{\mu}_{\rm F435W}= -2.5\,\log I_s(\phi)_{\rm F435W}+ 16.150,
\end{equation}

where $I_s(\phi)_{\rm F814W}$ and $I_s(\phi)_{\rm F435W}$ represent
the azimuthally averaged, background subtracted intensity (in counts
per second per pixel) in F814W and F435W respectively.  Because
the galaxies completely fill the field of view of the ACS/HRC and
WFPC2/PC detectors, a nominal sky background, as listed in the ACS
(Pavlosky et al. 2004a) or WFPC2 (Heyer et al. 2004) Instrument
Handbooks, was adopted.

\subsection{Dust obscuration and optical depth map}
\label{sec:dust}

Correction for dust obscuration is necessary to recover the intrinsic
stellar luminosity density and the spatial distribution of the ionized
gas within each galaxy.  We adopted the procedure described in
Ferrarese et al. (2006), which works under the assumption that the
F435W-F814W intrinsic color of the galaxy in the regions obscured by
dust can be estimated by linearly interpolating across the dust area
(or extrapolating, if the dust affects the center) the
azimuthally-averaged color measured in the regions unaffected by dust.
The dust absorption at each pixel then follows by comparing the
intrinsic and observed (extincted) color maps, once the ratio of the
absorption in F435W and F814W is known (see equation 2 of Ferrarese et
al. 2006). The latter, $A_{{\rm F435W}}/A_{{\rm F814W}}=2.2814$ (where
$A_{\lambda}$ is given in magnitudes), has been derived using the
extinction law of Cardelli et al. (1989), integrated over the filter
passbands, and assuming a ratio of total-to-selective absorption
$R_V=3.1$ (see also Schlegel et al. 1998). It is further assumed that
the dust lies in the galaxy foreground (the ``screen'' approximation);
the values quoted below for the dust absorption therefore represent
firm lower limits to the true absorption. Once the absorption in a
specific passband is known, the intrinsic fluxes are easily recovered
simply by correcting the measured fluxes for the derived magnitude
loss due to dust obscuration.

In producing the F435W-F814W color maps, the ACS images were rotated
(using the information about the position angle of the telescope axis
contained in the header) and resampled using {\tt PYDRIZZLE} in order
to match the orientation and scale of the lower-resolution WFPC2
images.

Dust absorption maps are shown in the left panel of
Figures~\ref{ha1836}--\ref{ha3565}.  In Abell 1836-BCG, the average
extinction (corrected to the $V-$band\footnotemark) within the regular
nuclear dust disk is $\langle A(V)\rangle=0.27$ mag, with values as
high as $A(V)_{max}= 0.85$ mag (in the screen approximation).  The
nuclear dust disk in Abell 3565-BCG is quite similar, with
$A(V)_{max}= 1.1$ mag and $\langle A(V) \rangle=0.21$ mag.  Recovering
a dust extinction map for Abell 2052-BCG, in which the dust is
distributed in an irregular filamentary pattern and the amount of
extinction is more modest, was more challenging. Within the inner
$\approx$ 1\hbox{${}^{\prime\prime}$}, we find $A(V)_{max}= 0.35$ mag
and $\langle A(V) \rangle=0.14$ mag. At larger radii, dust patches
with lower levels of dust obscuration, although visible in the color
maps, could not be properly modeled. These areas, which affect less
than 50\% of pixels along any given isophote, were simply masked when
deriving the galaxy's surface brightness profile.

The extinction corrected F814W surface brightness profiles of the sample
galaxies are shown in Figures~\ref{phot1836}--\ref{phot3565}.

\footnotetext{Extinction in all relevant passbands are related to A(V)
  as $A_{{\rm F435W}}/A_{{\rm V}}=1.362$, $A_{{\rm F814W}}/A_{{\rm
  V}}=0.597$, $A_{{\rm FR656N}}/A_{{\rm V}}=0.781,0.783$, and 0.805
  for Abell 1836-BCG, Abell 2052-BCG, and Abell 3565-BCG, respectively
  (Cardelli et al. 1989; Schlegel et al. 1998).}

\subsection{Deprojected stellar density models}
\label{sec:mge}

The deprojected luminosity density, $\Gamma (x,y,z)$, needed to
constrain the stellar gravitational potential, was recovered from the
redder (and thereby less affected by dust obscuration), extinction
corrected, WFPC2/F814W images, using the multi-Gaussian expansion
(MGE) method (Monnet at al. 1992; Emsellem et al. 1994), as
implemented by Cappellari (2002)\footnote{The code is publicly
available at http://www.strw.leidenuniv.nl/$\sim$mcappell/idl/}.

Briefly, the MGE procedure starts by determining the galaxy's average
ellipticity, position angle, and coordinates of the
luminosity-weighted center. The galaxy image is then divided in four
quadrants, and a number of photometric profiles is measured along
sectors uniformly spaced in angle from the major axis to the minor
axis. Surface brightness profiles from sectors in the four quadrants
are averaged together (an acceptable simplification for our galaxies
since position angle and isophotal center do not show a significant
radial dependency), and each is then fitted as the sum of Gaussian
components. The best-fitting MGE model surface brightness is then
determined iteratively by comparison with the observed surface
brightness, after having been convolved with the instrumental PSF
(generated, in our case, using the {\tt TinyTIM} package, Krist \&
Hook 1999, and parametrized as the sum of circular Gaussians).  The
Gaussian width coefficients of the MGE model were constrained to be a
set of logarithmically-spaced values, thus simplifying the fitting
algorithm into a general non-negative least-squares problem for the
corresponding Gaussian amplitudes.  During the entire procedure,
pixels for which correction for dust obscuration did not perform
correctly (as in the case of Abell 2052-BCG) were masked.

The best-fitting MGE model (prior to PSF convolution) is then
deprojected to a luminosity density $\Gamma(r)$ under the assumption
that the galaxy is spherically symmetric -- a justifiable
simplification since the isophotes are nearly circular throughout the
main body of our galaxies, including the radial range where the
ionized-gas kinematics probes the galaxy potential (where however
uncertainties are larger due to the presence of dust residuals, see
Figures~\ref{phot1836}--\ref{phot3565}).  For a radially-independent
mass-to-light ratio, \mlstar , the stellar mass density $\rho(r)$ is
then simply expressed as $\rho(r)=$ \mlstar $\Gamma(r)$, and the
stellar gravitational potential can be expressed in terms of error
functions. The best-fitting, PSF-convolved MGE models to the observed
surface-brightness profiles are shown in
Figures~\ref{mge1836}--\ref{mge3565}, along with the deprojected
luminosity density profiles (prior to PSF-convolution), and the
corresponding circular velocity curves [for \mlstar $ = 1$].

\subsection{Ionized-gas distribution}
\label{sec:hamap}

\begin{figure}
\epsscale{1.}
\plotone{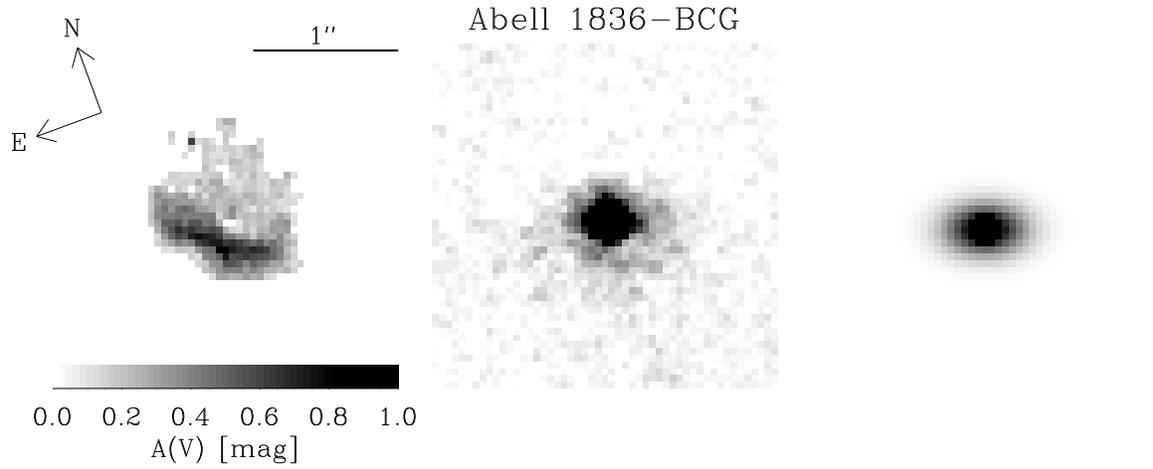}
\caption{{\small Optical depth map of Abell~1836-BCG ({\em left
panel}), and continuum-subtracted emission-line images in the
ACS/FR656N band pass  before ({\em central panel}) and
after deconvolution ({\em right panel}). The field of view is $6\Sec9\times
6\Sec9$. Orientation and scale are given in the left panel and they
are kept the same for all the panels.}}
\label{ha1836}
\end{figure}

Continuum-free emission-line images for the sample galaxies were
obtained by subtracting the WFPC2/F814W images from the ACS/FR656N
images which isolate the spectral region characterized by the
redshifted H$\alpha$ and \niipg\ emission lines.  All images were
previously extinction corrected as described in \S~\ref{sec:dust},
after having been resampled and rotated in order to match the pixel
scale and orientation of the lower-resolution WFPC2/F814W images.
Before subtraction, the FR656N images were multiplied by a factor
equal to the mean ratio between the F814W and resampled FR656N images,
calculated in emission-free regions.  The resulting emission line maps
are shown in the central panels of Figures~\ref{ha1836}--\ref{ha3565}.
The intrinsic (prior to PSF broadening) surface-brightness
distribution of the ionized gas, a key input to our modeling, was
obtained independently using both the iterative method based on the
Lucy-Richardson algorithm (Richardson 1972; Lucy 1974) and the MGE
method already described in \S~\ref{sec:mge}. According to van den
Bosch \& Emsellem (1998) the two methods give consistent results,
however, one might be preferable to the other depending on the
specifics of the images (e.g., signal-to-noise ratio S/N, presence of
asymmetric features, etc..). Both methods employed the PSF generated
using {\tt TINY TIM} for the WFPC2/F814W images. The Lucy-Richardson
algorithm, as implemented in the {\tt IRAF} task {\tt LUCY}, was found
to give satisfactory results in the case of Abell~2052-BCG and
Abell~3565-BCG, for which the method converged after only
five and six iterations, respectively. In the case of Abell~1836-BCG,
for which the emission-line images have lower S/N, {\tt LUCY} produced
unacceptable noise amplification before convergence could be reached,
and the intrinsic emission-line distribution was therefore recovered
using the MGE method. We note that the latter failed to produce
acceptable results for Abell~2052-BCG and Abell~3565-BCG, for which
the distribution of the ionized gas is less symmetric, violating the
four-fold symmetry required by the MGE algorithm.  The deconvolved
surface-brightness distributions of the emission line gas are shown in
the right panels of Figures~\ref{ha1836}--\ref{ha3565}.

\section{Observations, data reduction and analysis: STIS spectroscopy}
\label{sec:spectroscopy}

The spectroscopic observations of the sample galaxies were carried out
with STIS between 2003 August 16 and 2004 February 21 as part of
program GO-9838. The G750M grating was used at the prime tilt
in combination with the 0\Sec1$\times52''$~
slit for Abell~1836-BCG and Abell~2052-BCG, and the
0\Sec2$\times52''$~ slit for Abell~3565-BCG. The slit width was
selected to ensure that the sphere of influence of the SBH estimated
to reside in each galaxy (based on the M$_{\bullet}-\sigma$ and
M$_{\bullet}-$M$_{B}$ relations) could be comfortably resolved, while
at the same time allowing for reasonable throughput. The G750M grating
covers the wavelength range from 6480 \AA~to 7050 \AA, which includes
the \ha, \niipg, and \siipg\ emission lines. The dispersion is 0.56
\AA\ pixel$^{-1}$, and the instrumental resolution is 0.83 \AA\ (FWHM,
corresponding to $\sigma_{\rm instr}\simeq16.0$ \kms\ at \niig)
for the 0\Sec1$\times52''$~ slit, and 1.5 \AA\ ($\sigma_{\rm
instr}\simeq30.5$ \kms\ ) for the 0\Sec2$\times52''$~ slit. The
spatial scale of the 1024$\times$1024 SITe CCD is 0\Sec05
pixel$^{-1}$.

For each target galaxy, spectra were taken at three different slit
positions, the first crossing the nucleus of the galaxy, and two
additional ones displaced by one slit width (0\Sec1 for Abell~1836-BCG
and Abell~2052-BCG, and 0\Sec2 for Abell~3565-BCG) on either side,
perpendicularly to the slit axis. All slits were oriented along the
photometric major axis.  In the case of Abell~3565-BCG, for which the
wider slit allowed for shorter exposure times, one spectrum was taken
at each slit position. For Abell~1836-BCG and Abell~2052-BCG two
spectra were taken at each location, the second shifted along the slit
axis by four pixels to facilitate removal of bad pixels and other CCD
defects. A 90 s tungsten lamp flat exposure was taken after each
galaxy spectrum, to correct for the fringing affecting the STIS CCD at
wavelengths longer than $\approx 7000$ \AA.  A log of the STIS
observations can be found in Table~\ref{tab:STISlog}.

The spectra were reduced using the {\tt IRAF/CALSTIS} reduction
pipeline maintained by STScI. The basic reduction steps included
overscan subtraction, bias subtraction, dark subtraction, flatfield
correction, wavelength calibration as well as correction for geometric
distortion. Correction for fringing was performed with the {\tt
IRAF/DEFRINGE} task. In the case of Abell3565-BCG, for which only one
spectrum is available at each slit location, cosmic ray events and hot
pixels were removed using the task {\tt LACOS\_SPEC} (van Dokkum
2001).  For the other two galaxies, the two spectra obtained at the
same slit position were aligned with {\tt IRAF/IMSHIFT} using the
center of the stellar-continuum radial profile as a reference, and
then combined using {\tt IRAF/IMCOMBINE}. Cosmic ray rejection was
performed with the {\tt CRREJECT} option in {\tt IMCOMBINE},
additionally, pixels flagged by running {\tt LACOS\_SPEC} separately
on each spectrum were also removed. The sky background was determined
from the combined, calibrated spectra between $15''$ and $30''$ from
the nucleus and then subtracted. Finally, individual spectra were
extracted every 0\Sec05 (corresponding to 1 pixel) up to $\approx$
0\Sec2, 0\Sec3, and 0\Sec7 from the center for Abell~1836-BCG,
Abell~2052-BCG, and Abell~3565-BCG, respectively.  At larger radii,
the decreasing intensity of the emission lines required binning in the
spatial direction (see Table~\ref{tab:STISkin} and
Figures~\ref{1836kin}--\ref{3565kin} for details) to obtain a S/N of
at least 10 at the peak of the \niig\ emission line in the combined
spectrum, a condition deemed necessary for an accurate measurement of
the line kinematics (\S 4.1).  The final spectra along the major and
offset axes of the sample galaxies are plotted in
Figures~\ref{1836kin}(b)--\ref{3565kin}(b).

The location of the slit relative to the galaxy center 
was checked using STIS images taken
after acquisition of the target galaxies. A series of ``synthetic''
slits were extracted from the acquisition images by averaging four
adjacent columns for Abell~1836-BCG and Abell~2052-BCG (corresponding
to a 0\Sec1-wide slit), and eight adjacent columns for Abell~3565-BCG
(corresponding to a 0\Sec2-wide slit) at the nominal slit
orientation. The radial brightness profiles in each synthetic slit
were then compared to the profiles obtained by collapsing the spectrum
at the nuclear location along the wavelength direction.  The best
match was determined by minimizing the $\chi^2$ of the difference
between the light profile of the spectrum and the light profile
extracted from the acquisition image. In all cases, the slits
were found to be all positioned at the nominal locations.

\subsection{Measurement of the emission lines}
\label{sec:lines}

 The ionized-gas kinematics were measured from the narrow \niig\
  line, the brightest of the lines detected in our spectra.  This line
  is preferred to \ha\ since the latter is more severely blended, and
  could be significantly affected by emission from circumnuclear
  star-forming regions (e.g., Verdoes Kleijn et al. 2000; Coccato et
  al. 2006).

  The emission lines' central wavelengths, FWHMs, and intensities were
  measured following Beifiori et al. (2008). The best-fitting Gaussian
  parameters were derived using a non-linear least-squares
  minimization based on the robust Levenberg-Marquardt method (e.g.,
  Press et al. 1992) implemented by Mor\'e et al. (1980). The actual
  computation was done using the MPFIT algorithm\footnote{The updated
    version of this code is available on
    http://cow.physics.wisc.edu/$\sim$craigm/idl/idl.html} implemented
  by C.~B. Markwardt under the IDL\footnote{Interactive Data Language
    is distributed by Research System Inc.} environment.

  Because the line profiles are assumed to be Gaussian, it was often
  necessary to include two components in order to reproduce the
  extended wings observed for several of the emission lines (most
  notably \ha\ but, in some cases, also \niipg\ and \siipg). Lines in
  the same doublet (\nii\ and \sii) were assumed to have the same
  velocity and velocity width. A flux ratio of 1:2.967 was assumed for
  the two \nii\ lines, as dictated by atomic physics (e.g., Osterbrock
  1989).  The stellar continuum was approximated with a low-order
  polynomial.

  The major axis spectrum of Abell~1836-BCG displays both a narrow and
  a broad \ha\ line within the inner 0\Sec25. The broad-line emission
  appears to be spatially extended, with FWHM of 0\Sec17, compared to
  the typical FWHM of the STIS PSF ($\sim$ 0\Sec075).  Only the
  $-$0\Sec1 offset spectrum shows a broad \ha\ component within the
  inner 0\Sec1 (two rows), which possibly originates in the same
  nebular complex that gives rise to the broad-line emission observed
  along the central slit.

  Only the major axis spectrum of Abell 3565-BCG displays broad-line
  emissions.  Within the inner 0\Sec15 the spectrum shows an \ha\
  component while from 0\Sec15 to 0\Sec41 it shows broad components in
  \ha , \niipg , and \siipg. The broad \ha\ emission is spatially
  extended, with a FWHM of 0\Sec13.

  The major-axis and the $-$0\Sec1 offset spectra of Abell~2052-BCG
  show both narrow and broad components in \ha , \niipg , and \siipg\
  within the inner 0\Sec30. The 0\Sec1 offset spectrum shows all the
  broad components in the inner 0\Sec1.  The broad lines have a FWHM
  of 0\Sec10 for \ha, 0\Sec12 for the \nii\ doublet, 0\Sec18 for
  \siip, and 0\Sec08 for \siig\ (\sii\ broad lines have the same radial
  extension).  They appear to be less extended than for the other two
  galaxies, although still marginally resolved.

Figures~\ref{1836spec}--\ref{3565spec} show the results of the
multi-Gaussian fitting for the spectra extracted at the nuclear
location, and at a location displaced, along the major axis, by
$\approx 70$ parsec (0\Sec15, 0\Sec10, and 0\Sec25 for Abell 1836-BCG,
Abell 2053-BCG, and Abell 3565-BCG respectively).

Line-of-sight heliocentric velocities,
velocity dispersions and line intensities measured from the \niig\
line along the major and offset axes of the nuclear dust disks are
plotted in Figures~\ref{1836kin}(c)--\ref{3565kin}(c) and listed in
Table~\ref{tab:STISkin}.

\subsection{Ionized-gas kinematics and dust morphology}
\label{sec:kinematics}

In the remainder of this Section, we will comment briefly on the
kinematics of each individual galaxy. Before doing so, we note that
both Abell~1836-BCG and Abell~3565-BCG display regular and symmetric
velocity fields, as suggested by the smoothness and regularity of the
ionized-gas disks. In contrast, Abell~2052-BCG, which is characterized
by a more irregular emission-line morphology, shows irregular
kinematics.

\noindent {\bf Abell 1836-BCG}.  The rotation curve, velocity
  dispersion profile, and flux profile along the major axis are
  relatively regular (Figure~\ref{1836kin}c), which makes this object
  appealing for dynamical modelling. The rotation velocity increases
  rapidly within the inner 0\Sec1, reaching approximately 200 \kms,
  and flattening at larger distances.  The velocity dispersion peaks
  at about 370 \kms\ at the nuclear location, and declines on either
  side, although not symmetrically. The off-axis kinematics show a
  similar pattern.

\noindent {\bf Abell~2052-BCG}. The dust distribution in this galaxy
is somewhat irregular, with several filaments distributed in a
spiral-like pattern (Figure~\ref{acs2052}). This lack of symmetry is
reflected in the flux distribution of the ionized gas, which is also
not symmetric with respect to the center (lower panels of
Figure~\ref{2052kin}c), and in the \niig\ kinematics, which could be
measured out to about 0\Sec3 from the center along each slit positions
(Figure~\ref{2052kin}c). At the nuclear location, the maximum rotation
velocity (200 \kms) is reached at 0\Sec05, with the velocity
decreasing on either side, while the velocity dispersion curve is more
symmetric, reaching $\approx 500$
\kms at the nuclear location. The spectra flanking the nuclear spectra are
characterized by similar velocity dispersion profiles, but very
different rotation curves; in particular, almost no
rotation is observed along the slit displaced by 0\Sec1 to the
South-East. Because of this lack of regularity in the gas kinematics, 
we are only able to place an upper limit on the SBH mass of Abell~2052-BCG
(\S \ref{sec:upper_limit}).

\noindent {\bf Abell~3565-BCG.}   The \niig\ kinematics are measured
within 0\Sec8 from the center along all three slit locations
(Figure~\ref{3565kin}(c)).  
  Although the smooth and well defined
  dust disk in Abell~3565-BCG starts to tilt slightly at large radii
  from the center (Figure~\ref{acs3565}), such a warp is clearly
  visible, and presumably kinematically important, only beyond the
  region where the ionized-gas kinematics is measured.  The major-axis
  rotation velocity has a steep gradient, increasing to approximately
  250 \kms\ within 0\Sec1. It remains approximately constant out to
  the last observed point along the approaching side, while it
  decreases to about 100 \kms\ for radii larger than 0\Sec5 along the
  receding side.  The velocity dispersion profile is fairly symmetric
  and strongly peaked, reaching $\approx 450$ \kms\ at the center.
  The offset axes show rather similar kinematics, although the line
  fluxes are rather different.  The rotation velocity increases almost
  linearly with radius, up to $\approx 200$ \kms\ at $\approx$ 0\Sec3
  from the center, while the velocity dispersion remains fairly
  constant at 120 \kms. Like Abell 1836-BCG, Abell~3565-BCG is a
  promising candidate for dynamical modeling.

\section{Observations, data reduction and analysis: Ground-based spectroscopy}
\label{sec:groundspectrospcopy} 

Although not used in the dynamical analysis, large scale stellar
kinematics are necessary to place measured SBH masses on the
\mbh$-$\sigmac\ relation. Velocity dispersions measurements exist in
the literature for Abell~2052-BCG and Abell~3565-BCG, but not for
Abell~1836-BCG. Ground-based spectroscopic observations of
Abell~1836-BCG were therefore carried out in service mode with the
3.5-m New Technology Telescope (NTT) at the European Southern
Observatory (ESO) in La Silla (Chile) on 2007 May 20 and July 10 [DDT
programme ID 279.B-5004(A), P.I. E. Dalla Bont\`a]. The NTT mounted
the ESO Multi-Mode Instrument (EMMI) in red medium-dispersion
spectroscopic (REMD) mode, using the grating No. 7 with 600 $\rm
grooves\,mm^{-1}$ in first order with a 1.0 arcsec $\times$ 5.5 arcmin
slit. The detector was the No. 62 MIT/LL CCD with $2048\,\times\,
4096$ pixels of $15\,\times\,15$ $\rm \mu m^2$. After a $2\times2$
on-chip pixel binning it yielded a wavelength coverage between about
4546 \AA\ and 6096 \AA\ with a reciprocal dispersion of 0.83 $\rm
\AA\,pixel^{-1}$. The instrumental resolution was 2.5 \AA\ (FWHM)
corresponding to $\sigma_{\it inst}\approx55$ \kms\ at 5900 \AA. The
spatial scale was 0.332 arcsec pixel$^{-1}$. We took three spectra of
2400 s along the major axis of the galaxy (PA$=54 \dg$ ). The
integration time was split into three exposures to deal with cosmic
rays. Spectra of six giant stars with spectral type ranging from G7III
to K5III were obtained, to use as templates. Arc lamp spectra were
taken before and after every exposure to allow an accurate wavelength
calibration.  The average seeing FWHM during the observing runs was
1.0 arcsec in May 20 and 1.1 arcsec in July 10, as measured by the ESO
Differential Image Motion Monitor.

All spectra were bias subtracted, flatfield corrected, cleaned of
cosmic rays, corrected for bad pixels and columns, and wavelength
calibrated using standard {\tt IRAF} routines. It was checked that the
wavelength rebinning was done properly by measuring the difference
between the measured and predicted wavelengths (Osterbrock et al.
1996) for the brightest night-sky emission lines in the observed
spectral ranges.The spectra were co-added using the center of the
stellar continuum as reference.  The contribution of the sky was
determined from the outermost $\sim30$ arcsec and $\sim60$ arcsec at
the two edges of the resulting spectrum, respectively, and then
subtracted. A one-dimensional sky-subtracted spectrum was obtained for
each kinematical template star. Flux calibration was not performed.

The stellar velocity dispersion was measured from the galaxy
absorption features present in the wavelength range including the Mg
line triplet ($\lambda\lambda\,5164,5173,5184$ \AA) using the Fourier
Correlation Quotient method (Bender 1990) as done in Corsini et
al. (2007).  HR7429 (K3III) was adopted as the kinematical template.
The stellar velocity dispersion was measured within an aperture of
$1/8 r_e$. It was then applied the correction of Jorgensen et
al. (1995) to derive the velocity dispersion within a circular
aperture of radius $1/8 r_e$. The correction was negligible, because
the stellar velocity dispersion profile shows constant values, within
the errors, toward the center.  The stellar velocity dispersion of
Abell 1836-BCG within a circular aperture of size $1/8 r_e$
(Table~\ref{tab:sample}) resulted to be $\sigma_c=309\pm11$ \kms.

  A consistent value ($\sigma_c=308\pm10$ \kms) was obtained by
  measuring the velocity dispersion with the Penalized Pixel-Fitting
  Method (Cappellari \& Emsellem 2004) as done in M\'endez-Abreu et
  al. (2008).
  The measurement of the velocity dispersion from spectra obtained in
  the Mg triplet region can be problematic for massive ellipticals,
  due to possible mismatch of the abundance ratios with those of the
  adopted stellar templates (e.g., Barth et al. 2003). To address this
  issue, the Mg region was masked out (Fig.~\ref{fig:sigma}). A lower
  value of $\sigma_c=288\pm9$ \kms\ was found after applying the
  aperture correction, and this was adopted for Abell 1836-BCG.

\section{Dynamical modeling}
\label{sec:model}

Modeling of gas dynamics from HST data has steadily improved since the
first successful applications of the method (Harms et al. 1994;
Ferrarese et al. 1996; Barth et al. 2001; Sarzi et al. 2001; Marconi
et al. 2003). For this study, we will use the procedure described in
Coccato et al. (2006), where a detailed description of the method can
be found. Briefly, a synthetic velocity field is generated assuming
that the ionized gas is moving in circular orbits under the combined
gravitational potential of stars and SBH. The gas is assumed to be
confined in an infinitesimally thin disk centered at the photometric
center of the galaxy.  The model is projected onto the plane of the
sky for a given inclination of the gaseous disk, and then degraded to
simulate the actual setup of the spectroscopic observations.  The
latter step includes accounting for the width and location (namely
position angle and offset with respect to the center) of each slit,
instrumental PSF and charge bleeding between adjacent pixels. The free
parameters of the model are the mass \mbh\ of the SBH, the
mass-to-light ratio \mlstar\ of the stellar component (which we give
in the $I-$band), and the inclination $i$ of the gaseous disk; both
\mlstar\ and $i$ are assumed to be radially invariant.  Although $i$
can be estimated from the minor-to-major axial ratio of the disk,
which is easily measured from the images, previous studies (Ferrarese
et al. 1996, 1999; Sarzi et al. 2001; Shapiro et al. 2006) have shown
that slight warps are common. When the ionized-gas emission arises
predominantly from the innermost region of the dust disk, it is
therefore best to treat $i$ as a free parameter.  The location of the
slits, as well as the surface brightness distribution of the ionized
gas (derived as described in \S~\ref{sec:hamap}) are treated as input.
\mbh, \mlstar, and $i$ are determined by finding the model parameters
which produce the best match to the observed velocity curve, obtained
by minimizing $\chi^2 = \sum (v - v_{\it mod})^2/\delta^2(v)$ where
$v\pm\delta(v)$ and $v_{\it mod}$ are the observed and model velocity
along the different slit positions, respectively.

It has been noted in a number of studies of the kinematics of gas
disks (Ferrarese et al. 1996, 1998; van der Marel \& van den Bosch
1998; Verdoes Kleijn et al. 2000, 2002, 2007) that the measured
velocity dispersion of the emission lines greatly exceeds what is
expected from instrumental broadening and radial variations in
rotational velocity of the gas within a single slit element, and
requires an intrinsic component of random motion. In our model, this
is assumed to be described by a radial function of the form $\sigma(r)
= \sigma_0 + \sigma_1 e^{-r/r_\sigma}$. In principle, the exact
characterization of $\sigma(r)$ affects the recovered velocity field,
and therefore $\sigma_0$, $\sigma_1$, and $r_\sigma$ should be treated
as free parameters in the model, along with \mbh\ , $i$, and
\mlstar. In practice, however, previous studies (Verdoes Kleijn et
al. 2000; Barth et al. 2001; Coccato et al. 2006) have found that
$\sigma_0$, $\sigma_1$, and $r_\sigma$ are fairly insensitive to the
values of \mbh\ , $i$, and \mlstar. 

  We have run a set of preliminary models in order to find the shape
  of the intrinsic velocity dispersion profile that is required to
  match our data, within the typical range of our key model
  parameters. Specifically, we explored a reasonable grid of values
  for black hole mass, inclination, and stellar $I-$band mass-to-light
  ratio, finding for each combination of these parameters the values
  of $\sigma_0$, $\sigma_1$, and $r_\sigma$ that minimized $\chi^2 =
  \sum (v - v_{\it mod})^2/\delta^2(v) + \sum (\sigma - \sigma_{\it
    mod})^2/\delta^2(\sigma)$ where $\sigma\pm\delta (\sigma)$ and
  $\sigma_{\it mod}$ are the observed and the model velocity
  dispersion along the major axis. In our final modeling procedure we
  have then fixed $\sigma_0$, $\sigma_1$, and $r_\sigma$ to the values
  that were found for the best of such preliminary models (see next
  Sections).

In the following, the results for each galaxy will be discussed separately.

\subsection{Abell~1836-BCG}

We explored a three-dimensional grid of models with $0 \leq M_\bullet
\leq 3.6\times10^{10}$ \msun~ in 2.4$\times10^8$ \msun\ steps, $
0^\circ \leq i \leq 90^\circ$ in $1^\circ$ steps, and $ 0\leq
(M/L)_\star \leq 33$ \mlsun~ in 0.4 \mlsun\ steps. Within this range,
the intrinsic velocity dispersion of the gas best able to reproduce
the observables is $\sigma(r) = 12 + 204 \,e^{-r/0.1 {\rm~kpc}}$ km
s$^{-1}$.  The model adopts three parameters to fit 28 data points
(the values of the rotational velocity), for a total of 25 degrees of
freedom. The best model fitting the observed rotation curve (given the
intrinsic velocity dispersion profile given above) requires
$M_\bullet=3.61^{+0.41}_{-0.50}\times 10^9$ \msun,
$i={69.0\dg}^{+1.7\dg}_{-1.9\dg}$, and \mlstar $={5.0}^{+1.8}_{-0.9}$
\mlsun (in the $I-$band), where the errors on $M_\bullet$, $i$, and
\mlstar, are quoted at the $1\sigma$ confidence level. The model,
which has $\chi^2 = 85.3$ and a reduced $\chi^2_r=3.4$, is compared to
the observed spectrum and kinematics in Figure~\ref{mod2d1836} and
Figure~\ref{obs_mod1836}, respectively.

The small-scale asymmetries present in the model rotation curve (most
evident at the central position) are a direct consequence of using in
the models the intrinsic, PSF-deconvolved H$\alpha$+[NII] intensity
map (bottom panel of Figure~\ref{obs_mod1836}).  The observed rotation
curves are well reproduced in general; the slight mismatch observed at
$r \sim 0\Sec1$ for the slit located on the North-West side (-0\Sec1)
of the nucleus (Figure~\ref{obs_mod1836}) might be due to the presence
of a slight warp in the outer part of the gas disk, which is not
reproducible within the model's framework.  

The inclination angle derived from modeling the kinematic data
($i={69.0\dg}^{+1.7\dg}_{-1.9\dg}$), which extend only as far as $r
\lesssim 0\Sec2$, is significantly different from the inclination
angle suggested by the axial ratio of the outer edge ($r \approx
0\Sec5$) of the disk, $i= 46^\circ\pm5^\circ$. This suggests the
presence of a warped structure, although such a tilting does not need
to extend to the innermost regions where the kinematics was measured,
and effectively our images can not constrain the gas geometry.

Figures~\ref{contour1_1836}--\ref{contour3_1836} show the $1\sigma$,
$2\sigma$ and $3\sigma$ confidence levels (following the
$\Delta\chi^2$ variations expected for two free parameters, i.e.,
2.30, 6.17 and 11.8, Press et al. 1992) in the two-dimensional space
of two of the fitted parameters, where the third parameter is held
fixed at the best fitted value listed above.  Finally,
Figure~\ref{sigma_est1836} shows $1\sigma$, $2\sigma$, and $3\sigma$
confidence levels individually on \mbh, $i$, and \mlstar, according to
the $\Delta\chi^2$ variations expected for one parameter (i.e., 1, 4,
and 9, Press et al. 1992), marginalizing over all other parameters.

\subsection{Abell~3565-BCG}

\noindent

The rotation curves along the three slit positions, for a total of 90
velocity data-points, were fitted for a grid of model parameters
defined by $0 \leq M_\bullet \leq 6.4\times10^{9}$ \msun~ in
2.6$\times10^7$ \msun\ steps, $ 0^\circ \leq i \leq 90^\circ$ in
$1^\circ$ steps, and $ 0\leq (M/L)_\star \leq 12$ \mlsun\ in 0.3
\mlsun\ steps. Within this range, the intrinsic velocity dispersion of
the gas best able to reproduce the observables is $\sigma(r) = 122 +
98\, e^{-r/0.01 {\rm~kpc}}$ km s$^{-1}$.  The best model, which has
$\chi^2 = 1518$ and a reduced $\chi^2_r=17.5$ (for 87 degrees of
freedoms), requires $M_\bullet=1.34^{+0.21}_{-0.19}\times 10^9$ \msun,
$i={66.0\dg}^{+3.5\dg}_{-3.4\dg}$ and \mlstar $=6.3^{+1.1}_{-1.0}$
\mlsun, where all errors are given at the $1\sigma$ confidence level.
 Such formal uncertainties were conservatively estimated after
  scaling up our errors on the velocity measurement until $\chi^2_r=1$
  (following Sarzi et al. 2001), acknowledging that our thin-disk
  model fails to reproduce some of the velocity structure observed in
  particular at the central and South-Eastern slit positions. Contrary
  to the case of Abell 1836-BCG, the dynamical inclination angle of
  the disk is much closer to the morphological inclination angle, $i=
  71^\circ\pm1^\circ$, derived from the axial ratio of the outer edge
  of the disk ($r\approx1\Sec4$).  Although our kinematic data extend
almost to this radius ($r\lesssim 0\Sec8$) in practice the luminosity
weighting assigns more leverage to the data from the innermost
region. The model is compared to the observed spectrum and kinematics
in Figure~\ref{mod2d3565} and Figure~\ref{obs_mod3565}, respectively.

Figures~\ref{contour1_3565}--\ref{contour3_3565} show the $1\sigma$,
$2\sigma$, and $3\sigma$ confidence levels in the two-dimensional
space of two of the fitted parameters, where the third parameter is
held fixed at the best fitted value listed
above. Figure~\ref{sigma_est3565} shows $1\sigma$, $2\sigma$, and
$3\sigma$ confidence levels on \mbh, $i$, and \mlstar\ alone,
marginalizing over all other parameters. It is worth noticing that
mass-to-light ratios derived by the models refer to the nucleus of the
galaxy, and might differ from the stellar population characterizing
the galaxy on larger scale.

\subsection{Abell~2052-BCG }
\label{sec:upper_limit}

In the case of Abell~2052-BCG, the asymmetric nature of the gas
kinematics (\S 4.2) prevents the application of the method described
above to pinpoint \mbh.  An upper limit on the SBH mass can however
still be obtained as described in Sarzi et al. (2002).  In brief, an
upper limit can be derived under the assumption that the central
\niig\ emission-line width, $\sigma$, is produced by gas moving on
circular orbits in a coplanar, randomly oriented disk in the Keplerian
potential of the central black hole.  If the line broadening is caused
in part by non-gravitational forces, the derived upper-limit would
well exceed the actual SBH mass.

The model was implemented under the assumption that the intrinsic
radial flux profile of the gas can be described by a Gaussian:
\begin{equation}\label{flux}
F(r)=F_0+F_1\,\exp[-(r-r_0)^2/2\sigma_{flux}^2],
\end{equation}
where the parameters $F_0$, $F_1$, $r_0$, and $\sigma_{flux}$ are
derived by matching $F(r)$, after convolution with the STIS PSF, to
the observed emission flux profile along the slit.  For a given
black-hole mass, and under the previous assumptions regarding the gas
geometry and kinematics, the model produces two-dimensional projected
maps for the moments of the line-of-sight velocity distribution
(LOSVD) at any position $(x,y)$ on the sky:
\begin{equation}
 \overline{\Sigma v^k}(x,y)=\int {\rm LOSVD} (x,y,v_z) v_z^k {\rm
d}v_z\,\, (k=0,1,2).
\end{equation}
$\overline{\Sigma v^k}$ is then convolved with the STIS PSF; finally,
the velocity dispersion, $\sigma$, which corresponds to
$(\overline{\Sigma v^2}/\overline{\Sigma v^0} - (\overline{\Sigma
v^1}/\overline{\Sigma v^0})^2 )^{1/2}$, is extracted within a
$0.05''\times0.1''$ aperture for comparison to the observables.

There is no information on the orientation of the gaseous disk within
the central aperture, nor is it possible to estimate it by analyzing
the dust morphology, which is irregular and does not show a disk-like
structure.
In principle, a lower-limit on the SBH mass can be estimated
assuming an edge configuration, whereas a model with a perfectly
face-on disk would allow for an infinite SBH mass.  Fortunately
such extreme configurations are statistically rare, and considering
that randomly oriented disks have uniformly distributed cos$\,i$ we
can estimate 1$\sigma$ upper and lower limits on \mbh\ by using models
with $i=33^{\circ}$ and $i=81^{\circ}$, corresponding to cos$\,i = 0.84$
and 0.16, respectively.
For each $i$, the SBH mass was varied in 0.1$\times10^6$ \msun\
steps between $6\times10^6 \leq M_\bullet \leq 11.0\times10^9$ \msun~
until the observed gas velocity dispersion was matched.
The models were run for three values of mass-to-light ratio:
$(M/L)_\star$=0, 5.0 \mlsun\ (derived for Abell
1836-BCG), and 6.3 (as in the case of Abell 3565-BCG) \mlsun.  The
most conservative upper limit is obtained by neglecting altogether the
stellar contribution to the gravitational potential, \mlstar=0,
giving \mbh $\leq 4.6\times 10^9$ \msun\ and \mbh $\leq 1.8\times 10^9$
\msun\ for $i= 33\dg$ and $i= 81\dg$ respectively. Consistent with
Sarzi et al. (2002), adopting \mlstar=5.0 or 6.3 \mlsun, has very
little effect: \mbh\ decreases slightly to $3.6\times 10^9$ \msun\ and
$3.3\times 10^9$ \msun\ respectively for $i = 33\dg$ and $9.8\times
10^8$ \msun\ and $7.7\times 10^8$ \msun\ respectively for $i =
81\dg$. Figure~\ref{upper_limit} shows the dependence of \mbh\ on the
assumed \mlstar\ for $i=33\dg$ and $i=81\dg$.

\section{Discussion and conclusions}
\label{sec:conclusions}

We have presented a dynamical analysis aimed at constraining the
masses of the SBHs in three brightest
cluster galaxies: Abell 1836-BCG, at a distance of 147.2 Mpc, Abell
2052-BCG, at a distance of 141.0 Mpc, and Abell 3565-BCG, at a distance
of 50.8 Mpc. The models are based on data obtained with the Hubble Space
Telescope. Broad-band WFPC2 and broad and narrow-band ACS images were
used to constrain the luminosity profile and the distribution of the
ionized gas, as traced by the H$\alpha+$[N~{\small II}] emission, 
while STIS was employed to measure the rotation
velocity and velocity dispersion profiles, from the [N~{\small II}]$
\,\lambda6583$ emission, along three parallel slits positions, the
first aligned along the photometric major axis, and the others
adjacent to the first on either side of the nucleus.

In the case of Abell~1836-BCG and Abell~3565-BCG, the regular
morphology and kinematics observed for the ionized gas led to secure
black hole mass detections of \mbh$=$ $3.61^{+0.41}_{-0.50}\times
10^9$ \msun\ and $1.34^{+0.21}_{-0.19}\times 10^9$ \msun\ (where the
uncertainties represent 1$\sigma$ errors\footnote{Barth et
al. 2001 show that the total error budget for gas-dynamical
modelling is likely to be twice as large as such formal error
estimates.}), respectively. For Abell-2052-BCG, which displays
irregular kinematics, an upper limit of \mbh$\leq 4.60 \times10^9$
\msun\ was derived under the conservative assumption of a negligible
stellar contribution to the gravitational potential, and an
inclination angle for the gas disk of $33\dg$.
At face value, the black hole in Abell~1836-BCG is the most massive to
have been dynamically measured to-date.

In our modeling we have neglected the potential impact on our SBH mass
determination of assigning a dynamical origin to the additional
kinematic broadening that we have used in our models to match the
observed profile for the gas velocity dispersion. Although it has been
argued that this additional turbulence does not affect the bulk motion
of the gas (e.g. van der Marel \& van den Bosch 1998), it is important
to notice that ignoring the possibility that the gas might be
supported by dynamical pressure will lead to {\it under-}estimate the
SBH mass. For instance, Barth et al. (2001) have shown that including
a classical asymmetric drift correction led to an 12\% increase of
their best \mbh\ value.
Unfortunately, we could not repeat this analysis in our models since
the observed line-broadening exceeds by far the circular velocity,
breaking down the epicyclic approximation on which the asymmetric
drift correction is based. Nonetheless, our best SBH mass measurements
are still likely to be underestimations of the real SBH mass.

Figure~\ref{scal_rel} shows the location of Abell~1836-BCG,
Abell~2052-BCG, and Abell~3565-BCG in the \mbh$-$\sigmac\ (Ferrarese \&
Ford 2005; Tremaine et al. 2002) and near-infrared \mbh$-$\lbulge\
(Graham 2007; Marconi \& Hunt 2003) planes. $K$-band Two-Micron
All-Sky Survey (2MASS) magnitudes were retrieved from the NASA/IPAC
Infrared Science Archive, and corrected for Galactic extinction
following Schlegel et al. (1998).  Stellar velocity dispersions are
from this paper (see \S \ref{sec:groundspectrospcopy}) for Abell
1836-BCG, from Smith et al. (2000) in the case of Abell~3565-BCG (we
adopt his high S/N measurement, $\sigma=335\pm 12$ \kms), and Tonry
(1985) for Abell~2052-BCG. We note that published estimates of
$\sigma$ for Abell 2052-BCG range between $250$ \kms\ and $370$ \kms;
the high end of this spread is probably due to contamination from a
companion galaxy located only $8''$ away in the North-East direction.
The contribution from this galaxy was explicitly accounted for by
Tonry (1985), whose value, $\sigma=253\pm 12 $ \kms, we have adopted.
Correcting the adopted values of $\sigma$ to a circular aperture of
size $1/8 r_e$ (Table~\ref{tab:sample}), following Jorgensen et
al. (1995), we find $\sigma_c=288 \pm 9$ \kms, $\sigma_c=233 \pm 11$
\kms, and $\sigma_c=322\pm12$ \kms\ for Abell~1836-BCG, Abell~2052-BCG
and Abell~3565-BCG, respectively.

The SBH mass detected in Abell~3565-BCG, \mbh =
$1.34^{+0.21}_{-0.19}\times 10^9$ \msun\ , is consistent both with the
\mbh$-$\sigmac\ (Ferrarese \& Ford 2005) and the $K$-band
\mbh$-$\lbulge\ (Graham 2007) relations, which predict \mbh =
$1.7^{+2.0}_{-0.9}\times 10^9$ \msun\ (the error is computed adopting
a 0.34 dex scatter in \mbh) and \mbh = $1.1^{+1.1}_{-0.6}\times 10^9$
\msun\ (adopting a 0.30 dex scatter), respectively.

For Abell~2052-BCG, although the conservative upper limit on the black
hole mass obtained for reasonable assumptions for \mlstar\ and $i$ (\S
5.3), is consistent with both the \mbh$-$\sigmac\ (Ferrarese \& Ford
2005) and the $K$-band \mbh$-$\lbulge\ (Graham 2007) relations, these
predict significantly different masses, $3.5^{+4.1}_{-1.9}\times 10^8$
\msun\ (adopting a 0.34 dex scatter) and $1.3^{+1.3}_{-0.6}\times10^9$
\msun\ (adopting a 0.30 dex scatter) respectively (but note the above
caveat regarding the measurement of $\sigma$ for this
galaxy). Although the sense of the discrepancy is consistent with the
one argued for by Lauer et al. (2007), the lack of a precise
determination of the SBH mass in this galaxy prevents us from
establishing which, if either, of the two relations better predicts
$M_{\bullet}$ in this case.  The only two BCGs in the literature with
a dynamically measured SBH mass are M87 (Macchetto et al. 1997) and
NGC 1399 (Houghton et al. 2006). M87 appears to have a more massive
black hole than expected based on its $K-$band luminosity while it obeys
the \mbh$-$\sigmac\ relation. NGC 1399 is consistent with both the
relations within their scatter.

Abell 1836-BCG appears to be an outlier in all SBH scaling
relations. The dynamically measured SBH mass reported in this paper,
$M_\bullet=3.61^{+0.41}_{-0.50}\times 10^9 $ \msun, is larger, at the
3$\sigma$ level, than the value of \mbh = $9.6^{+9.5}_{-4.8}\times
10^8$ \msun\ (adopting a 0.30 dex scatter), predicted by the $K-$band
\mbh$-$\lbulge\ relation (Graham 2007). It is also larger, at the
3$\sigma$ level, than the value of \mbh = $9.77^{+11.6}_{-5.30}\times
10^8$ \msun\ (adopting a 0.34 dex scatter), predicted by the
\mbh$-$\sigmac\ relation of Ferrarese \& Ford (2005).
Finally, the SBH mass of Abell 1836-BCG is not consistent
with the value of $M_\bullet=6.3^{+11.5}_{-4.1}\times 10^8 $ \msun\
predicted by the fundamental plane relation for SBHs by Hopkins et
al. (2007), adopting an effective radius $r_e=13\Sec1=9.3$ kpc with a
scatter of 0.45 dex in \mbh.

The existence of extremely massive SBHs as outliers to the empirical
\mbh$-$\sigmac\ relation would not be surprising. As shown by Lauer et
al. (2007), the \mbh$-$\lbulge\ relation predicts a greater abundance
of the most massive SBHs, given the luminosities of the BCGs and the
lack of galaxies with velocity dispersions larger than $\sim 350$
\kms. At the same time, the observed luminosity function of AGN
implies the presence of SBHs as massive as $\sim 5\times 10^9$~\msun\
within the distance of the clusters we have observed, and SBHs mergers
can increase the highest SBH masses in these massive clusters up to
$\sim 10^{10}\,$\msun\ (Yoo et al. 2007). A larger sample of SBH masses
measured in massive clusters would be required to test models of the
effect of mergers in increasing the largest SBH masses in the
universe.

In conclusion, both the \mbh$-$\sigmac\ and
\mbh$-$\lbulge\ relations appear at their high-\mbh\ end to be
consistent with the SBH mass measured for one BCG, Abell 3565-BCG, but
inconsistent with another one, Abell 1836-BCG. For the remaining
target, Abell 2052-BCG, although the ionized-gas kinematics allowed us
only to set an upper-limit on \mbh\ it seems unlikely that this galaxy
could obey both relations simultaneously. The fact that Abell 1836-BCG
is an outliers in both the \mbh$-$\sigmac\ and \mbh$-$\lbulge\
relations would appear to weaken the claim of Lauer et al (2007) that
\lbulge\ is more reliable predictor of \mbh\ for BCGs. Overall, our results
might indicate that the scatter of SBH scaling relations increases at the
high end, although additional data are necessary to test this claim.

\acknowledgments { EDB acknowledges the Fondazione ``Ing. Aldo
    Gini'' for a research fellowship, the Herzberg Institute of
    Astrophysics, Victoria, BC, and the University of Oxford, UK, for
    the hospitality while this paper was in progress. JM is supported
    in part by Spanish grants AYA 2006-06341 and AYA2006-15623-C02-01.
    Support for program GO-9838 was provided through a grant from the
    Space Telescope Science Institute, which is operated by the
    Association of Universities for Research in
    Astronomy, Inc., under NASA contract NAS5-26555.
    We thank V. Debattista, R. Trotta, and M. Cappellari for the
    stimulating discussions.}

\clearpage


\begin{table}
\begin{scriptsize}
\caption{{\small Basic parameters of the sample galaxies}\label{tab:sample}}
\begin{center}
\begin{tabular}{llcccrcrrc}
\tableline
\tableline
\multicolumn{1}{c}{Galaxy} &
\multicolumn{1}{c}{Other name} &
\multicolumn{1}{c}{Type} &
\multicolumn{1}{c}{$D_{25}$} &
\multicolumn{1}{c}{$r_{e}$} &
\multicolumn{1}{c}{m$_B$}&
\multicolumn{1}{c}{m$_K$} &
\multicolumn{1}{c}{$cz$} &
\multicolumn{1}{c}{D} &
\multicolumn{1}{c}{$\sigma_c$} \\
\multicolumn{1}{c}{} &
\multicolumn{1}{c}{} &
\multicolumn{1}{c}{[RC3]} &
\multicolumn{1}{c}{[arcmin]} &
\multicolumn{1}{c}{[arcsec]} &
\multicolumn{1}{c}{[mag]} &
\multicolumn{1}{c}{[mag]} &
\multicolumn{1}{c}{[km\,s$^{-1}$]} &
\multicolumn{1}{c}{[Mpc]} &
\multicolumn{1}{c}{[km\,s$^{-1}$]} \\
\multicolumn{1}{c}{(1)} &
\multicolumn{1}{c}{(2)} &
\multicolumn{1}{c}{(3)} &
\multicolumn{1}{c}{(4)} &
\multicolumn{1}{c}{(5)} &
\multicolumn{1}{c}{(6)} &
\multicolumn{1}{c}{(7)} &
\multicolumn{1}{c}{(8)} &
\multicolumn{1}{c}{(9)} &
\multicolumn{1}{c}{(10)} \\
\tableline
Abell~1836-BCG  & PGC~49940 & SA0$^-$: & 1.4 & $13.10^{\small {\rm a}}$ & $14.56\pm0.14^{\small {\rm a}}$ & $9.99\pm0.04$  & 11036 &  147.2 &  $288\pm9^{\small {\rm a}}$                        \\
Abell~2052-BCG  & UGC~9799  & E        & 1.8 & $55.98^{\small {\rm b}}$ & $14.4\pm0.03^{\small {\rm b}}$ & $9.55\pm0.06$   & 10575 &  141.0 & $233\pm11^{\small {\rm b}}$ \\
Abell~3565-BCG	& IC~4296   & E        & 3.4 & $41.41^{\small {\rm a}}$ & $11.61\pm0.05^{\small {\rm c}}$ & $7.50\pm0.02$   &  3834 &   50.8 & $322\pm12^{\small {\rm c}}$ \\
\tableline
\end{tabular}
\tablecomments{
Col. (1): Galaxy name from Abell (1958) and Abell et al. (1989). 
Col. (2): Alternative identification.
Col. (3): Morphological classification from RC3.
Col. (4): Apparent major isophotal diameter measured at the surface
          brightness level $\mu_B= 25$ mag arcsec$^{-2}$ from RC3.
Col. (5): Effective radius (a: from RC3, b: from Hudson et al. 2001).
Col. (6): Total apparent $B-$band magnitude (a, c: from RC3, b: from
Postman \& Lauer 1995).
Col. (7): Total apparent $K-$band magnitude from 2MASS.
Col (8): Systemic velocities in the frame of the Cosmic Microwave Background, from Laine et al. (2003)
Col. (9): Distances from systemic velocity and $H_0=75$ 
          $\mathrm{km\;s}^{-1}\mathrm{\;Mpc}^{-1}$, except for Abell
3565-BCG (Jensen et al. 2003).
Col. (10): Stellar velocity dispersion (a: this paper; b: Smith et al. 2000; c: Tonry 1985) corrected to a circular aperture of size $1/8 r_e$.}
\end{center}
\end{scriptsize}
\end{table}

\clearpage

\begin{table}
\begin{scriptsize}
\caption{{\small Log of the ACS and WFPC2 observations.}
\label{tab:ACSlog}}
\begin{center}
\begin{tabular}{ccccccc}
\tableline
\tableline
\multicolumn{1}{c}{Galaxy} &
\multicolumn{1}{c}{Filter} &
\multicolumn{1}{c}{Pivot $\lambda$} &
\multicolumn{1}{c}{Width} &
\multicolumn{1}{c}{Exp. Time} &
\multicolumn{1}{c}{Obs. Date} \\
\multicolumn{1}{c}{} &
\multicolumn{1}{c}{} &
\multicolumn{1}{c}{\AA} &
\multicolumn{1}{c}{\AA} &
\multicolumn{1}{c}{[s]} &
\multicolumn{1}{c}{} \\
\multicolumn{1}{c}{(1)} &
\multicolumn{1}{c}{(2)} &
\multicolumn{1}{c}{(3)} &
\multicolumn{1}{c}{(4)} &
\multicolumn{1}{c}{(5)} &
\multicolumn{1}{c}{(6)} \\
\tableline
\tableline
\multicolumn{6}{c}{ACS}\\
\tableline
Abell~1836-BCG  & F435W  & 4311   & 310    & $2\times895.0$       & 2004 Feb 20 \\
                & FR656N & 6803   & 136    & $2\times1382.5$     & 2004 Feb 20 \\
\tableline		  	   	    
Abell~2052-BCG  & F435W  & 4311   & 310    & $2\times900.0$       & 2003 Sep 3 \\
                & FR656N & 6788   & 136    & $2\times1380.0$     & 2003 Sep 3 \\
\tableline		  	   	    
Abell~3565-BCG  & F435W  & 4317   & 293    & $2\times900.0$       & 2004 Aug 7 \\
              	& FR656N & 6645   & 133    & $2\times1388.0$     & 2004 Aug 7 \\
\tableline
\tableline
\multicolumn{6}{c}{WFPC2}\\
\tableline
Abell~1836-BCG  & F814W  & 8012   & 1539   &  $2\times500.0$                             &  2001 Apr 11 \\
\tableline
Abell~2052-BCG  & F814W  & 8012   & 1539   &  $2\times500.0$                             &  2001 Apr 17 \\
\tableline
Abell~3565-BCG  & F814W  & 8012   & 1539   &$1\times900.0+5\times1100.0+4\times1300.0$ &  1996 Jan 19-20  \\
\tableline
\tableline
\end{tabular}
\tablecomments{
Col. (1): Galaxy name. 
Col. (2): Filter name.
Col. (3): Pivot wavelength (ACS: Sirianni et al. 2005; WFPC2: Heyer et
al. 2004).
Col. (4): Filter width (ACS: Sirianni et al. 2005; WFPC2: Heyer et
          al. 2004).
Col. (5): Total exposure time.
Col. (6): Observation date.}
\end{center}
\end{scriptsize}
\end{table}

\clearpage

\begin{table}
\caption{{\small Log of the STIS observations.}
\label{tab:STISlog}}
\begin{center}
\begin{tabular}{lrrcrr}
\tableline
\tableline
\multicolumn{1}{l}{Galaxy} &
\multicolumn{1}{c}{Offset} &
\multicolumn{1}{c}{PA} &
\multicolumn{1}{c}{Exp. Time} &
\multicolumn{1}{c}{Obs. Date} \\
\multicolumn{1}{c}{} &
\multicolumn{1}{c}{[$''$]} &
\multicolumn{1}{c}{[$^\circ$]} &
\multicolumn{1}{c}{[s]} &
\multicolumn{1}{c}{} \\
\multicolumn{1}{c}{(1)} &
\multicolumn{1}{c}{(2)} &
\multicolumn{1}{c}{(3)} &
\multicolumn{1}{c}{(4)} &
\multicolumn{1}{c}{(5)} &\\
\tableline
                &   0.0   & $-127.9$ & $2040+2800$    & 2004  Feb 21 \\ 
Abell~1836-BCG  & $-0.1$  & $-127.9$ & $2\times 2800$ & 2004  Feb 21 \\	
                & $+0.1$  & $-127.9$ & $2\times 2800$ & 2004  Feb 21 \\	
\tableline			         	       			    
                &   0.0   &   $51.0$ & $ 2030+2800$   & 2003  Sept 2 \\ 
Abell~2052-BCG  & $-0.1$  &   $51.0$ & $2\times 2800$ & 2003  Sept 2 \\	
                & $+0.1$  &   $51.0$ & $2\times 2800$ & 2003  Sept 2 \\
\tableline			          	       			    
              	&   0.0   &   $75.4$ & $2010$         & 2003  Aug 16 \\
Abell~3565-BCG 	& $-0.2$  &   $75.4$ & $2735$         & 2003  Aug 16 \\
              	& $+0.2$  &   $75.4$ & $2735$         & 2003  Aug 16 \\
\tableline
\end{tabular}
\tablecomments{
Col. (1): Galaxy name. 
Col. (2): Measured offset of the slit center with respect to the
          galaxy nucleus.
Col. (3): Position angle of the slit, measured from north to east.
Col. (4): Total exposure time.
Col. (5): Observation date.}
\end{center}
\end{table}

\clearpage

\begin{deluxetable}{rccrccrcc}
\tabletypesize{\scriptsize}
\tablecaption{\niig\ kinematics.\label{tab:STISkin}}
\tablewidth{0pt}
\tablehead{
\colhead{$r$} & \colhead{$V$} & \colhead{$\sigma$} &\colhead{$r$} & \colhead{$V$} & \colhead{$\sigma$} &\colhead{$r$} & \colhead{$V$} & \colhead{$\sigma$} \\
\colhead{[$''$]} &\colhead{[km\,s$^{-1}$]} &\colhead{[km\,s$^{-1}$]} &\colhead{[$''$]} &\colhead{[km\,s$^{-1}$]} &\colhead{[km\,s$^{-1}$]} &\colhead{[$''$]} &\colhead{[km\,s$^{-1}$]} &\colhead{[km\,s$^{-1}$]} \\
\colhead{(1)} &\colhead{(2)} &\colhead{(3)} &\colhead{(4)} &\colhead{(5)} &\colhead{(6)} &\colhead{(7)} &\colhead{(8)} & \colhead{(9)}\\
}
\startdata
\multicolumn{9}{c}{Abell~1836-BCG}\\
\multicolumn{3}{c}{$+0\Sec1$ offset} &
\multicolumn{3}{c}{major axis} &
\multicolumn{3}{c}{$-0\Sec1$ offset} \\
\tableline
$ -0.23$& 11382$\pm$  25&   80   $\pm$    28&$-0.23 $&   11500  $\pm$   28&   165  $\pm$    27& $  -0.23   $&   11418  $\pm$  24  &  133$\pm$ 24 \\
$ -0.18$& 11398$\pm$  51&   256  $\pm$    52&$-0.18 $&   11422  $\pm$   31&   289  $\pm$    31& $  -0.18   $&   11361  $\pm$  17  &  137$\pm$ 18 \\
$ -0.13$& 11402$\pm$  53&   369  $\pm$    51&$-0.13 $&   11455  $\pm$   18&   266  $\pm$    22& $  -0.13   $&   11341  $\pm$  18  &  200$\pm$ 18 \\
$ -0.08$& 11288$\pm$  26&   306  $\pm$    27&$-0.08 $&   11413  $\pm$   11&   294  $\pm$    13& $  -0.08   $&   11283  $\pm$  18  &  275$\pm$ 19 \\
$ -0.03$& 11307$\pm$  19&   286  $\pm$    20&$-0.03 $&   11369  $\pm$   11&   370  $\pm$    13& $  -0.03   $&   11237  $\pm$  30  &  396$\pm$ 38 \\
$  0.02$& 11199$\pm$  17&   228  $\pm$    17&$ 0.02 $&   11118  $\pm$   13&   362  $\pm$    17& $   0.02   $&   11094  $\pm$  35  &  260$\pm$ 31 \\
$  0.07$& 11202$\pm$  43&   346  $\pm$    45&$ 0.07 $&   11001  $\pm$   13&   208  $\pm$    17& $   0.07   $&   11082  $\pm$  30  &  301$\pm$ 43 \\
$  0.12$& 11094$\pm$  21&   132  $\pm$    22&$ 0.12 $&   10993  $\pm$   11&   114  $\pm$    12& $   0.12   $&   11024  $\pm$  60  &  241$\pm$ 64 \\
$  0.17$& 11091$\pm$  28&   164  $\pm$    29&$ 0.17 $&   10978  $\pm$   20&   123  $\pm$    21&             &                     &              \\
$  0.22$& 11014$\pm$  14&   51   $\pm$    19&$ 0.22 $&   10982  $\pm$   21&   92   $\pm$    23&             &                     &              \\

\tableline
\tableline
\multicolumn{9}{c}{Abell~2052-BCG}\\
\multicolumn{3}{c}{$+0\Sec1$ offset} &
\multicolumn{3}{c}{major axis} &
\multicolumn{3}{c}{$-0\Sec1$ offset} \\
\tableline

$  -0.25$&  10189$\pm$ 27 &  103 $\pm$ 28 & $  -0.30  $& 10172$\pm$ 56  &  209$\pm$   56  & $  -0.30  $&  10100 $\pm$  17  &     83$\pm$    18 \\
$  -0.20$&  10220$\pm$ 39 &  229 $\pm$ 37 & $  -0.25  $& 10156$\pm$ 34  &  179$\pm$   31  & $  -0.25  $&  10011 $\pm$  33  &    213$\pm$    30 \\
$  -0.15$&  10247$\pm$ 17 &  158 $\pm$ 16 & $  -0.20  $& 10130$\pm$ 17  &  183$\pm$   17  & $  -0.20  $&  9979  $\pm$  13  &    193$\pm$    13 \\
$  -0.10$&  10318$\pm$ 16 &  169 $\pm$ 16 & $  -0.15  $& 10160$\pm$ 15  &  176$\pm$   23  & $  -0.15  $&  9912  $\pm$  11  &    102$\pm$    15 \\
$  -0.05$&  10330$\pm$  8 &  132 $\pm$  9 & $  -0.10  $& 10154$\pm$ 10  &  171$\pm$   16  & $  -0.10  $&  9908  $\pm$   7  &    186$\pm$     9 \\
$   0.00$&  10329$\pm$ 11 &  169 $\pm$ 12 & $  -0.05  $& 10220$\pm$  9  &  150$\pm$   13  & $  -0.05  $&  9930  $\pm$   7  &    199$\pm$     9 \\
$   0.0 $&  10348$\pm$ 15 &  211 $\pm$ 15 & $   0.00  $& 10436$\pm$ 18  &  294$\pm$   20  & $   0.00  $&  10075 $\pm$  12  &    168$\pm$    16 \\
$   0.10$&  10312$\pm$ 23 &  236 $\pm$ 22 & $   0.05  $& 10494$\pm$ 23  &  190$\pm$   28  & $   0.05  $&  10186 $\pm$  25  &    215$\pm$    29 \\
$   0.15$&  10311$\pm$ 16 &  176 $\pm$ 15 & $   0.10  $& 10364$\pm$ 19  &  187$\pm$   21  & $   0.10  $&  10242 $\pm$  26  &    162$\pm$    35 \\
$   0.20$&  10311$\pm$ 22 &  132 $\pm$ 22 & $   0.15  $& 10251$\pm$ 23  &  199$\pm$   21  & $   0.15  $&  10236 $\pm$  16  &    147$\pm$    16 \\
$   0.25$&  10313$\pm$ 21 &   97 $\pm$ 21 & $   0.20  $& 10251$\pm$ 16  &  133$\pm$   16  & $   0.20  $&  10242 $\pm$  21  &    191$\pm$    21 \\
$   0.30$&  10328$\pm$ 29 &  194 $\pm$ 29 & $   0.25  $& 10263$\pm$ 12  &  83 $\pm$   13  & $   0.25  $&  10296 $\pm$  15  &     73$\pm$    16 \\
$   0.36$&  10268$\pm$ 22 &   97 $\pm$ 23 & $   0.30  $& 10301$\pm$ 30  &  138$\pm$   30  & $         $&                   &                   \\
$   0.41$&  10459$\pm$ 35 &  109 $\pm$ 36 & $   0.36  $& 10293$\pm$ 88  &  159$\pm$   89  & $         $&                   &                   \\
$   0.46$&  10312$\pm$ 14 &   38 $\pm$ 19 & $   0.46  $& 10293$\pm$ 32  &  113$\pm$   33  & $         $&                   &                   \\
\tableline
\tableline
\multicolumn{9}{c}{Abell~3565-BCG}\\
\multicolumn{3}{c}{$+0\Sec1$ offset} &
\multicolumn{3}{c}{major axis} &
\multicolumn{3}{c}{$-0\Sec1$ offset} \\
\tableline

$ -0.76$&   4018$\pm$  20&   86$\pm$  23& $    -0.81 $&    4011 $\pm$    13   &    62$\pm$      16&  $   -0.71  $&      4015$\pm$       26 &      135   $\pm$  27\\
$ -0.71$&   4010$\pm$  11&   80$\pm$  12& $    -0.71 $&    4045 $\pm$    24   &   133$\pm$      25&  $   -0.61  $&      4054$\pm$       18 &      86    $\pm$  21\\
$ -0.66$&   4020$\pm$  11&   97$\pm$  12& $    -0.66 $&    4073 $\pm$    16   &   113$\pm$      17&  $   -0.56  $&      3999$\pm$       11 &      69    $\pm$  13\\
$ -0.61$&   4046$\pm$  12&  115$\pm$  13& $    -0.61 $&    4057 $\pm$    14   &    88$\pm$      16&  $   -0.51  $&      4006$\pm$       16 &      112   $\pm$  17\\
$ -0.56$&   4032$\pm$   9&  107$\pm$   9& $    -0.56 $&    4064 $\pm$    13   &   100$\pm$      14&  $   -0.46  $&      4017$\pm$       15 &      113   $\pm$  16\\
$ -0.51$&   4057$\pm$  10&  132$\pm$  11& $    -0.51 $&    4102 $\pm$    16   &   119$\pm$      17&  $   -0.41  $&      3987$\pm$       15 &      120   $\pm$  16\\
$ -0.46$&   4077$\pm$  12&  154$\pm$  12& $    -0.46 $&    4101 $\pm$    15   &   117$\pm$      16&  $   -0.36  $&      3976$\pm$       13 &      110   $\pm$  14\\
$ -0.41$&   4073$\pm$  11&  142$\pm$  12& $    -0.41 $&    4085 $\pm$    14   &   111$\pm$      15&  $   -0.30  $&      3970$\pm$       12 &      97    $\pm$  13\\
$ -0.36$&   4054$\pm$  13&  177$\pm$  12& $    -0.36 $&    4099 $\pm$    13   &   143$\pm$      14&  $   -0.25  $&      3954$\pm$        8 &      82    $\pm$  10\\
$ -0.30$&   4028$\pm$  12&  181$\pm$  12& $    -0.30 $&    4116 $\pm$    16   &   193$\pm$      16&  $   -0.20  $&      3939$\pm$        8 &      107   $\pm$   9\\
$ -0.25$&   4032$\pm$  13&  213$\pm$  14& $    -0.25 $&    4125 $\pm$    12   &   206$\pm$      12&  $   -0.15  $&      3898$\pm$        9 &      147   $\pm$  10\\
$ -0.20$&   4016$\pm$  10&  224$\pm$  10& $    -0.20 $&    4077 $\pm$    9    &   204$\pm$      10&  $   -0.10  $&      3866$\pm$        6 &      115   $\pm$   7\\
$ -0.15$&   3970$\pm$   8&  227$\pm$   9& $    -0.15 $&    4068 $\pm$    19   &   226$\pm$      14&  $   -0.05  $&      3833$\pm$        5 &      112   $\pm$   6\\
$ -0.15$&   3971$\pm$   8&  234$\pm$   8& $    -0.10 $&    4116 $\pm$    15   &   251$\pm$      15&  $    0.00  $&      3824$\pm$        4 &      109   $\pm$   4\\    
$ -0.10$&   3926$\pm$   5&  206$\pm$   5& $    -0.05 $&    4084 $\pm$    17   &   399$\pm$      18&  $    0.05  $&      3815$\pm$        4 &      97    $\pm$   4\\    
$ -0.05$&   3874$\pm$   4&  191$\pm$   4& $     0.00 $&    3830 $\pm$    13   &   451$\pm$      16&  $    0.10  $&      3804$\pm$        4 &      99    $\pm$   5\\    
$  0.00$&   3818$\pm$   3&  155$\pm$   3& $     0.05 $&    3673 $\pm$    6    &   254$\pm$       8&  $    0.15  $&      3746$\pm$       15 &     195    $\pm$  15\\    
$  0.05$&   3791$\pm$   3&  138$\pm$   3& $     0.10 $&    3607 $\pm$    6    &   203$\pm$       8&  $    0.20  $&      3725$\pm$       12 &     167    $\pm$  12\\    
$  0.10$&   3762$\pm$   5&  164$\pm$   5& $     0.15 $&    3590 $\pm$    8    &   171$\pm$      10&  $    0.25  $&      3681$\pm$       16 &     176    $\pm$  17\\
$  0.15$&   3729$\pm$   8&  201$\pm$   8& $     0.20 $&    3572 $\pm$   14    &   149$\pm$      15&  $    0.30  $&      3685$\pm$       16 &     134    $\pm$  16\\    
$  0.20$&   3686$\pm$   9&  204$\pm$   9& $     0.25 $&    3599 $\pm$   8     &   181$\pm$       8&  $    0.36  $&      3669$\pm$       17 &     155    $\pm$  17\\    
$  0.25$&   3655$\pm$  11&  165$\pm$  11& $     0.30 $&    3551 $\pm$   11    &   166$\pm$      11&  $    0.41  $&      3629$\pm$       22 &     163    $\pm$  23\\  
$  0.30$&   3652$\pm$  10&  150$\pm$  10& $     0.36 $&    3551 $\pm$   14    &   179$\pm$      14&  $    0.46  $&      3670$\pm$       19 &     168    $\pm$  20\\  
$  0.36$&   3648$\pm$   9&  130$\pm$   9& $     0.41 $&    3583 $\pm$   23    &   260$\pm$      23&  $    0.51  $&      3650$\pm$       19 &     130    $\pm$  20\\    
$  0.41$&   3636$\pm$   8&   96$\pm$   9& $     0.46 $&    3566 $\pm$   22    &   195$\pm$      23&  $    0.56  $&      3685$\pm$       11 &      81    $\pm$  12\\    
$  0.46$&   3621$\pm$  10&  120$\pm$  11& $     0.51 $&    3523 $\pm$   37    &   195$\pm$      37&  $    0.61  $&      3659$\pm$       13 &      99    $\pm$  15\\   
$  0.51$&   3588$\pm$  14&  140$\pm$  15& $     0.56 $&    3583 $\pm$   40    &   216$\pm$      41&  $    0.66  $&      3672$\pm$       21 &     150    $\pm$  22\\  
$  0.56$&   3582$\pm$  13&  149$\pm$  14& $     0.61 $&    3495 $\pm$   22    &   119$\pm$      23&  $    0.71  $&      3629$\pm$       17 &      85    $\pm$  19\\  
$  0.61$&   3595$\pm$  11&  109$\pm$  12& $     0.66 $&    3481 $\pm$   26    &    94$\pm$      29&  $    0.76  $&      3671$\pm$       16 &      68    $\pm$  20\\ 
$  0.66$&   3622$\pm$  13&  101$\pm$  14& $     0.76 $&    3552 $\pm$   29    &   119$\pm$      31&  $          $&                         &                     \\
$  0.71$&   3624$\pm$  18&   94$\pm$  20&             &                       &                   &              &                         &                     \\
\enddata
\tablecomments{
Col. (1)-(3): Galactocentric distance, line-of-sight heliocentric
  velocity (uncorrected for inclination), and line-of-sight velocity
  dispersion (uncorrected for instrumental velocity dispersion)
  measured from the [NII]$\lambda6583$ line in the $+0\Sec1$ offset spectrum.
Col. (4)-(6): As in Col. (1)-(3) but for the major-axis spectrum.
Col. (7)-(9): As in Col. (1)-(3) but for the $-0\Sec1$
  offset spectrum.}
\end{deluxetable}

\clearpage


\clearpage

\clearpage

\clearpage

\clearpage

\begin{figure}
\epsscale{1.}
\plotone{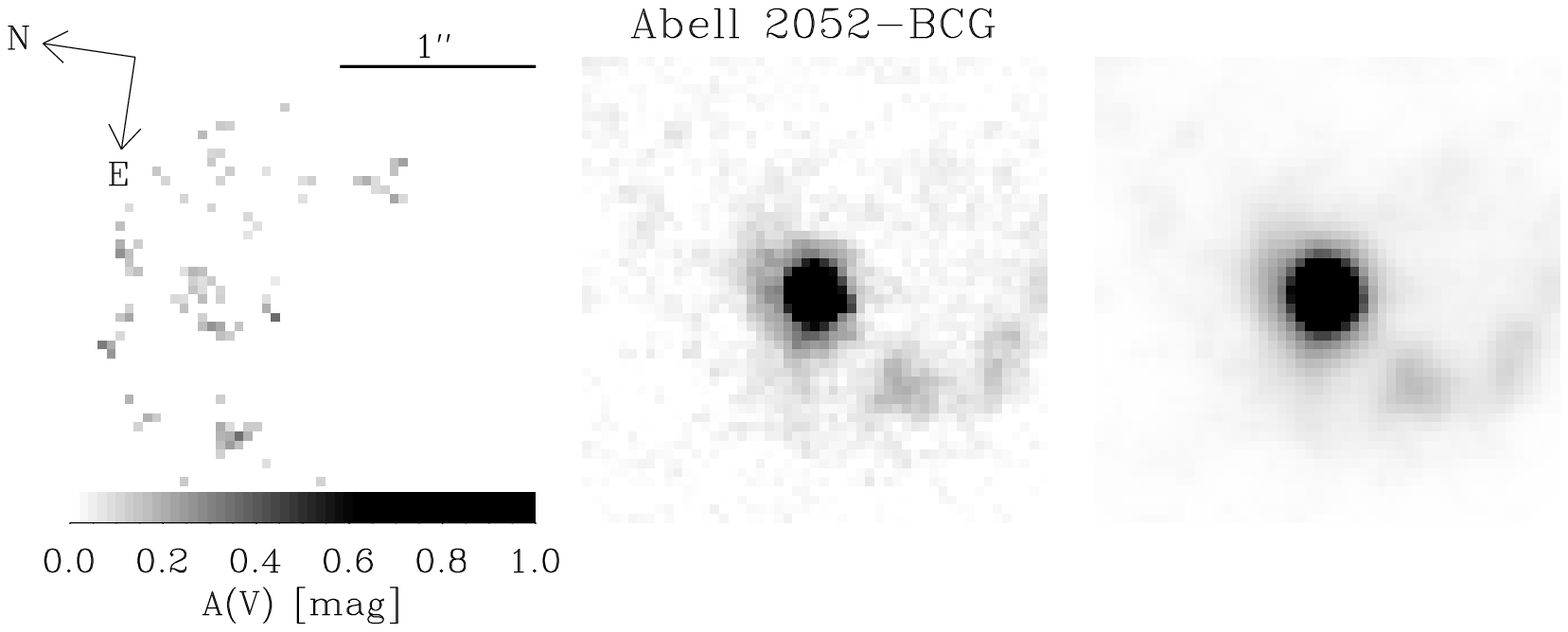}
\caption{{\small As in Figure~\ref{ha1836}, but for Abell~2052-BCG.}}
\label{ha3565}
\end{figure}

\begin{figure}
\epsscale{1.}
\plotone{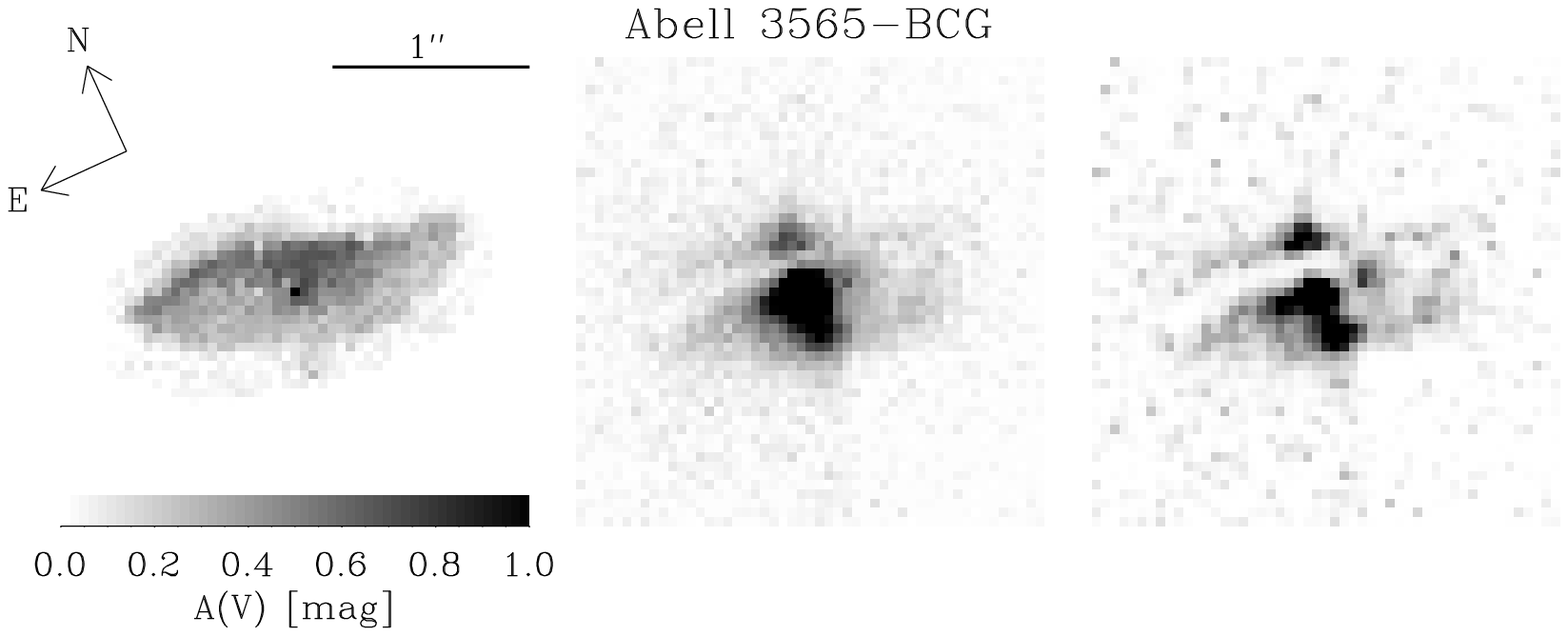}
\caption{{\small As in Figure~\ref{ha1836}, but for Abell~3565-BCG.}}
\label{ha3565}
\end{figure}

\clearpage

\begin{figure}
\epsscale{1.}
\plotone{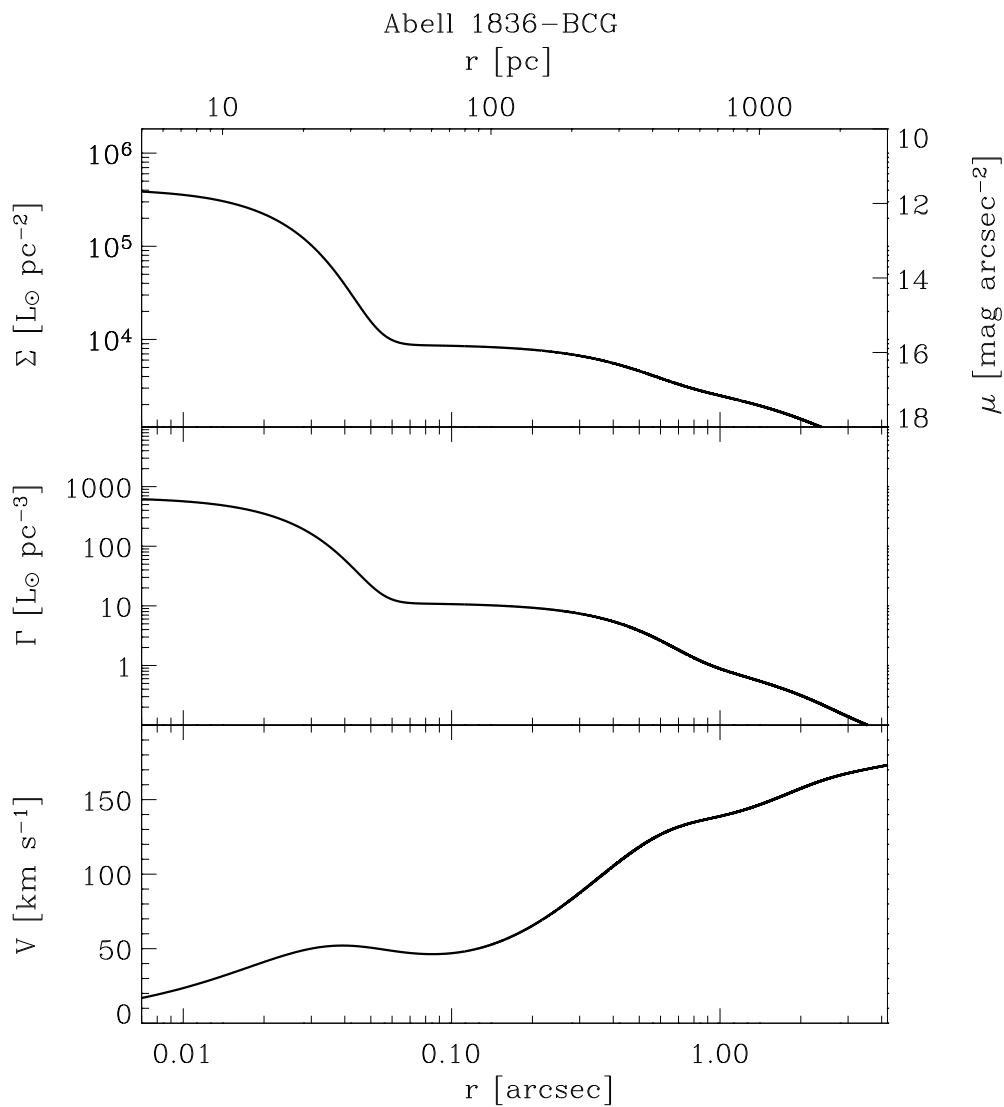}
\caption{{\small Deprojection steps for the stellar mass profile of
 Abell~1836-BCG. 
 {\em Top panel:} Multi-Gaussian fit to the PSF-deconvolved
 surface brightness profile derived from the extinction-corrected
 WFPC2/F814W image. {\em Middle panel:} Deprojected stellar luminosity
 density profile. {\em Bottom panel:} Circular velocity curve assuming \mlstar $ = 1$.}
\label{mge1836}}
\end{figure}

\begin{figure}
\epsscale{1.}
\plotone{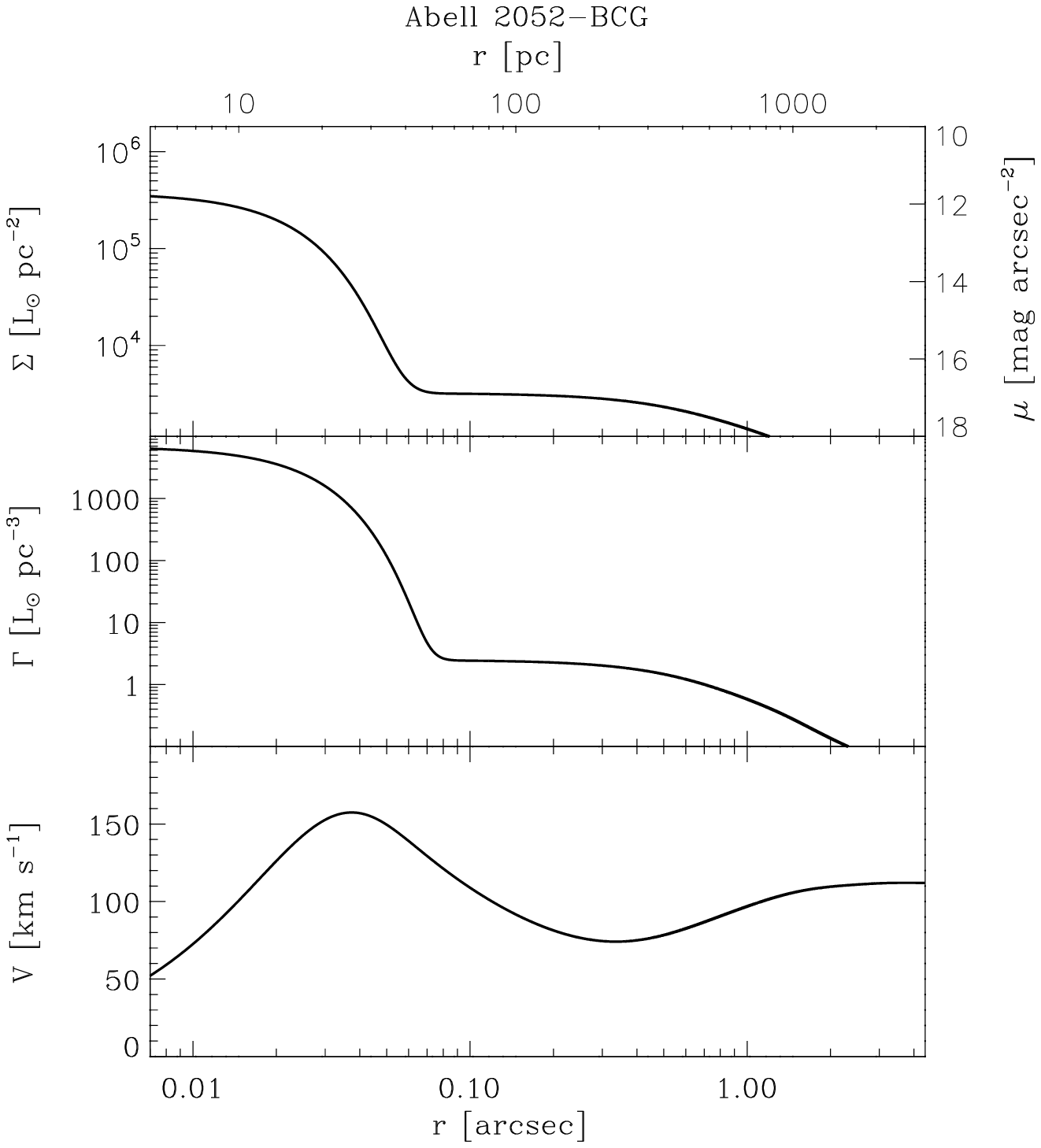}
\caption{{\small As in Figure~\ref{mge1836}, but for Abell~2052-BCG.}
\label{mge2052}}
\end{figure}

\begin{figure}
\epsscale{1.}
\plotone{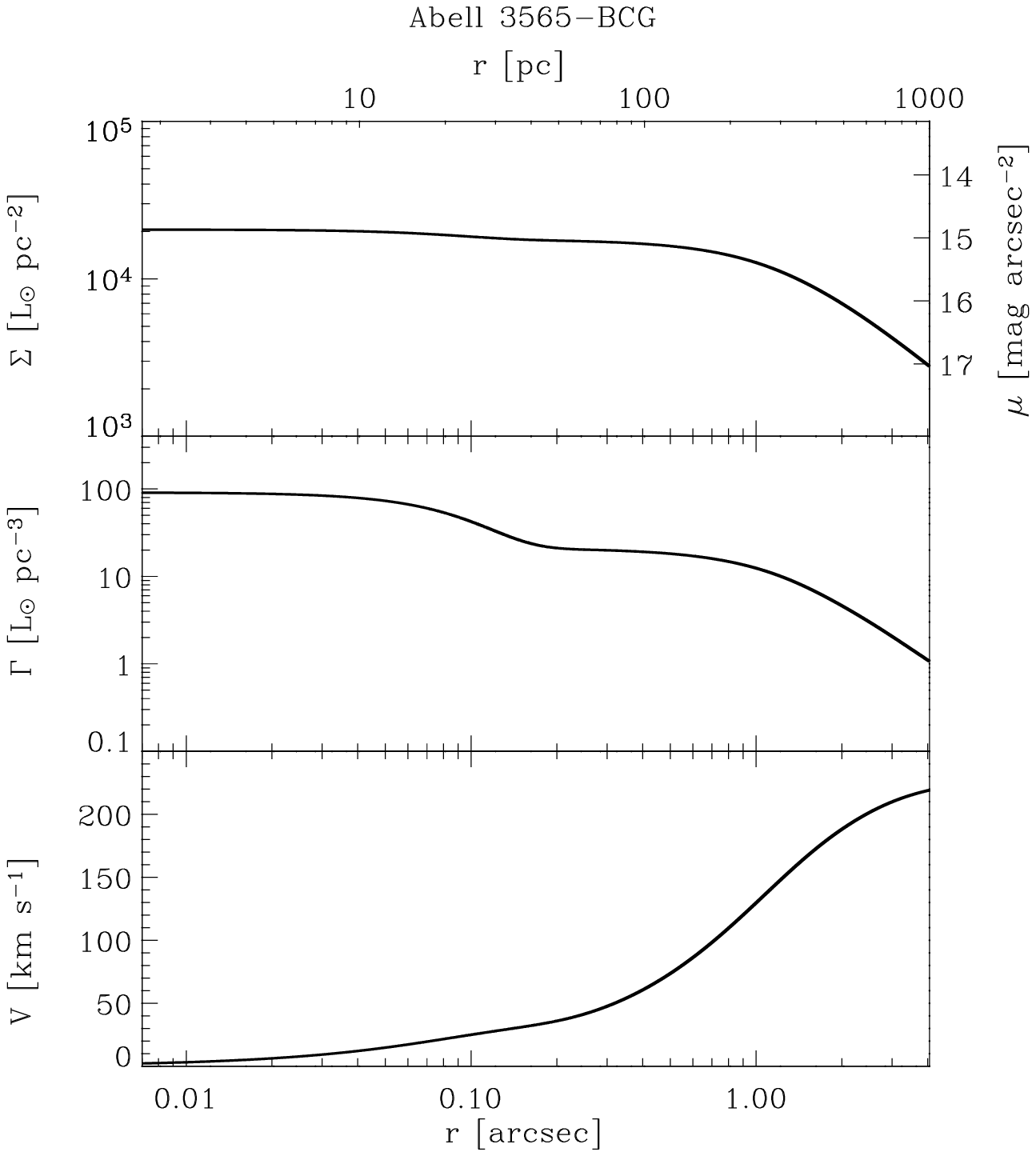}
\caption{{\small As in Figure~\ref{mge1836}, but for Abell~3565-BCG.}
\label{mge3565}}
\end{figure}

\clearpage

\begin{figure}
\epsscale{0.80}
\plotone{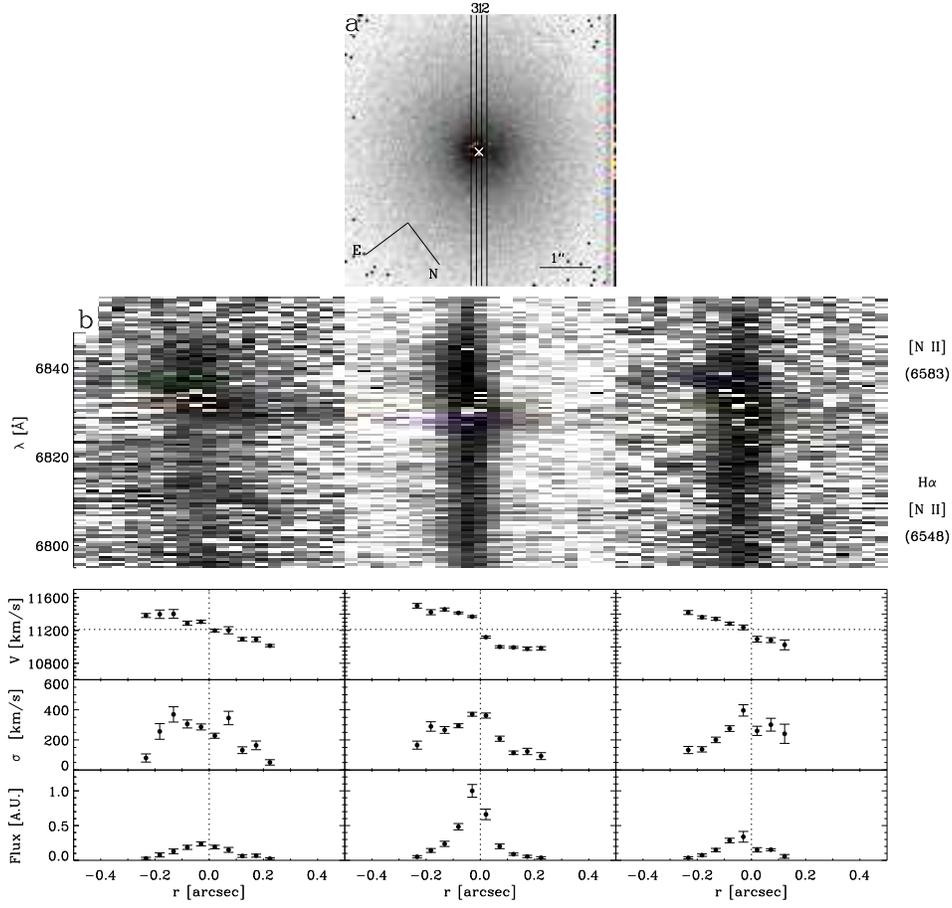}
\caption{
  {\small {\it a)\/} \hst\ STIS/F28X50LP acquisition image of
  Abell~1836-BCG.  The image has been rotated to the STIS instrumental
  frame. Orientation and scale are given. The white cross is the
  position of the nucleus from STIS acquisition procedure. The
  rectangles overplotted on the image show the actual locations of the
  slit during the spectroscopic observations.
  {\it b)\/} Portions of the bidimensional STIS spectra obtained in
  position 1 (major axis of the dust disk, {\it central panel\/}), 2
  ($-$0\Sec1 offset, {\it left panel\/}) and 3 ($+$0\Sec1 offset, {\it
  right panel\/}). The spectral region centered on the \ha\ emission
  line is shown after wavelength calibration, flux calibration and
  geometrical rectification. The spatial axis is horizontal and ranges
  between $-0\Sec5$ and $+0\Sec5$, while the wavelength axis is
  vertical and ranges from 6795 to 6856 \AA. Individual emission lines
  (\niip , \ha, and \niig) are identified on the right side of the
  figure.
  {\it c)\/} \niig\ kinematics from the spectra obtained in position 1
  ({\it central panels\/}), 2 ({\it right panels\/}) and 3 ({\it left
  panels\/}). For each slit position the line-of-sight velocity curve
  ({\it top panel\/}), the radial profile of the line-of-sight
  velocity dispersion (uncorrected for instrumental velocity
  dispersion, {\it middle panel\/}), and the radial profile of line
  flux in arbitrary units ({\it bottom panel\/}) are given.
  Heliocentric velocities and velocity dispersions are not corrected
  for inclination and instrumental velocity dispersion, respectively.}
\label{1836kin}}
\end{figure}
\clearpage

\begin{figure}
\epsscale{0.80}
\plotone{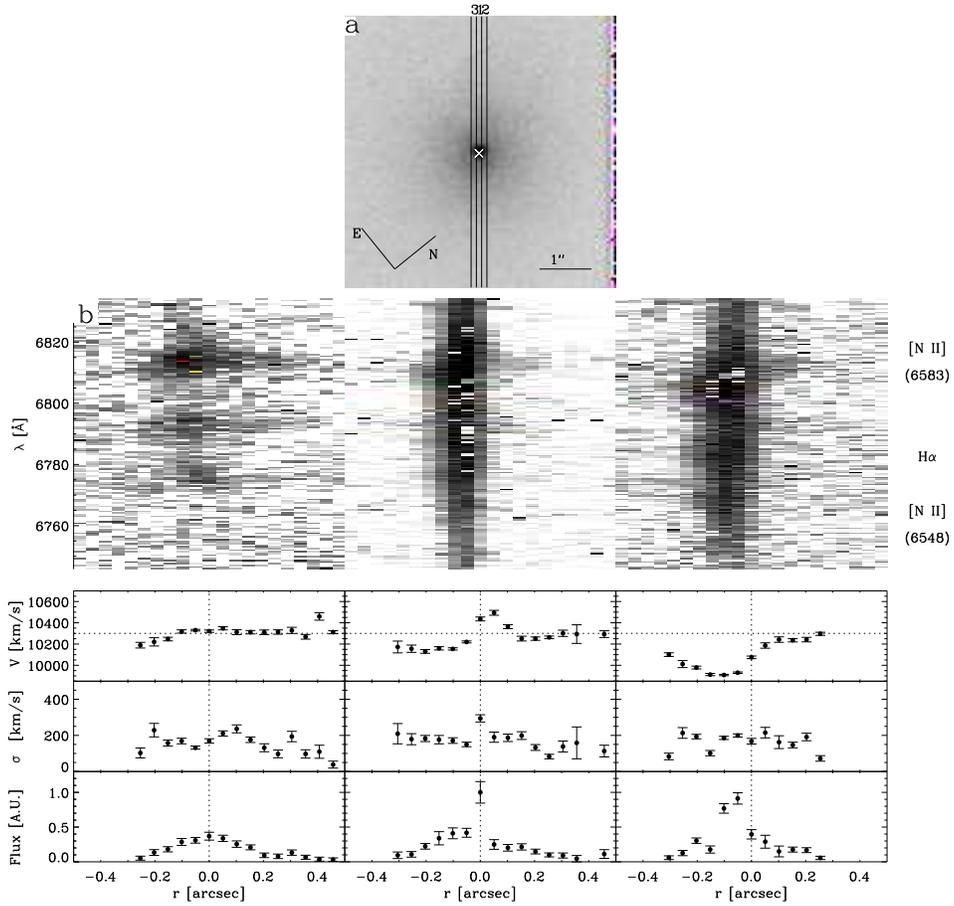}
\caption{
  {\small As in Figure~\ref{1836kin}, but for Abell~2052-BCG. In {\it
b)\/} the spatial axis is horizontal and ranges between $-0\Sec5$ and
$+0\Sec5$, while the wavelength axis ranges from 6746 to 6834 \AA.}  }
\label{2052kin}
\end{figure}
\clearpage

  \begin{figure}
  \epsscale{0.80}
  \plotone{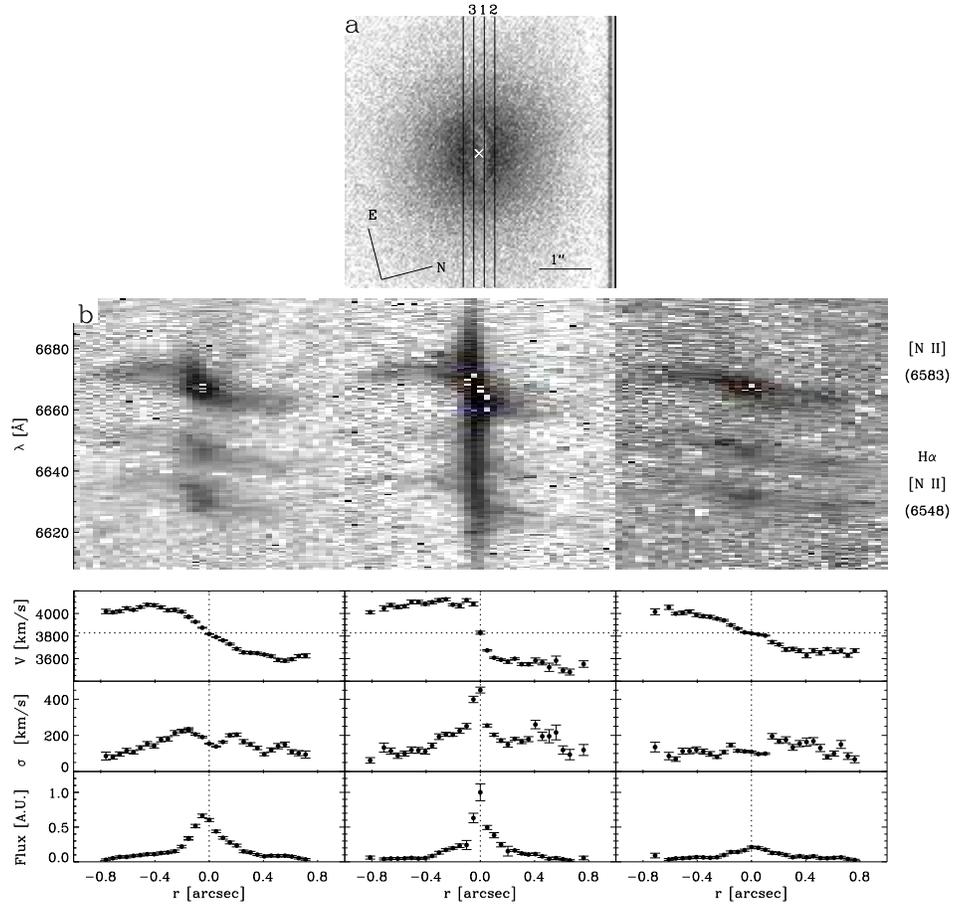}
  \caption{
    {\small As in Figure~\ref{1836kin}, but for Abell~3565-BCG. In
{\it b)\/} the spatial axis is horizontal and ranges between $-1\Sec0$
and $+1\Sec0$, while the wavelength axis ranges from 6608 to 6697
\AA. Positions 2 and 3 correspond to a slit-position offset of
$-$0\Sec2 and $+$0\Sec2, respectively.}}
  \label{3565kin}
  \end{figure}
  \clearpage

\begin{figure}
\epsscale{0.85}
\plotone{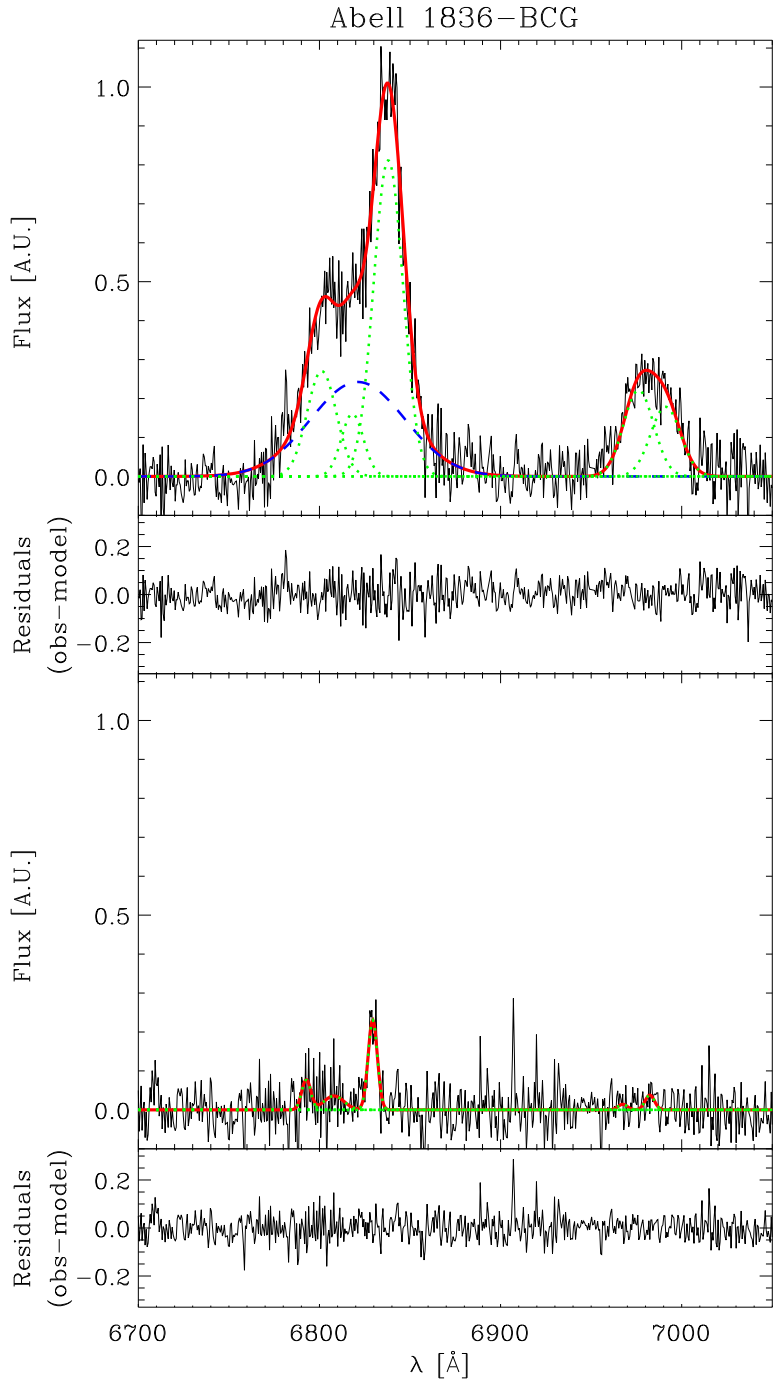}
\caption{
  {\small Nuclear spectrum ({\em upper panel\/}) and spectrum
    extracted at $r=+0\Sec15$ ({\em lower panel\/}), along the major
    axis of the dust disk of Abell~1836-BCG. Each spectrum is
      plotted with the Gaussian-model narrow (dotted lines) and broad
      (dashed lines) components, reconstructed blend, and residuals.}
\label{1836spec}}
\end{figure}

\begin{figure}
\epsscale{0.85}
\plotone{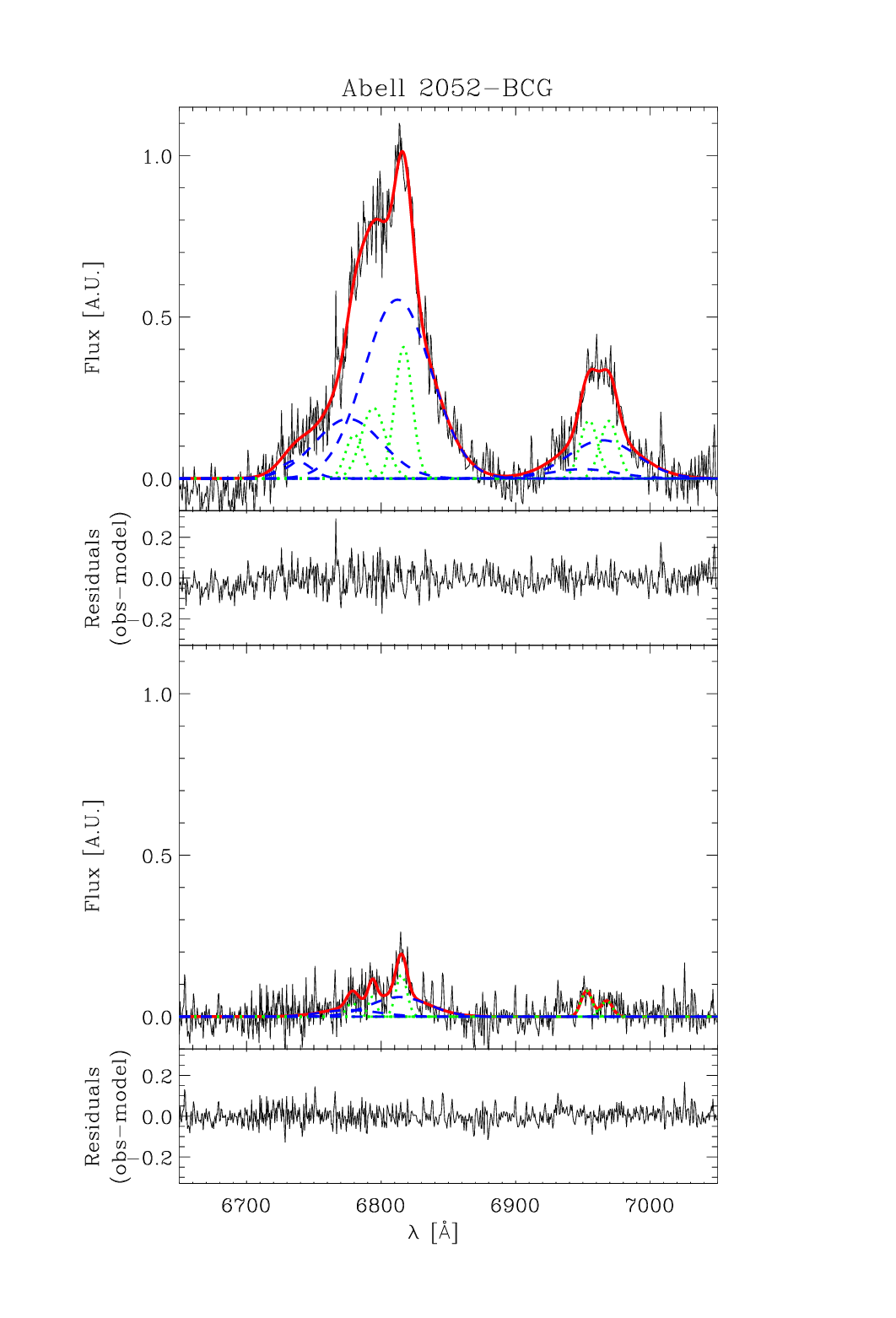}
\caption{{\small As in Figure~\ref{1836spec}, but for Abell~2052-BCG.
  The outer spectrum was extracted at $r=+0\Sec10$.}
\label{2052spec}}
\end{figure}

\begin{figure}
\epsscale{0.85}
\plotone{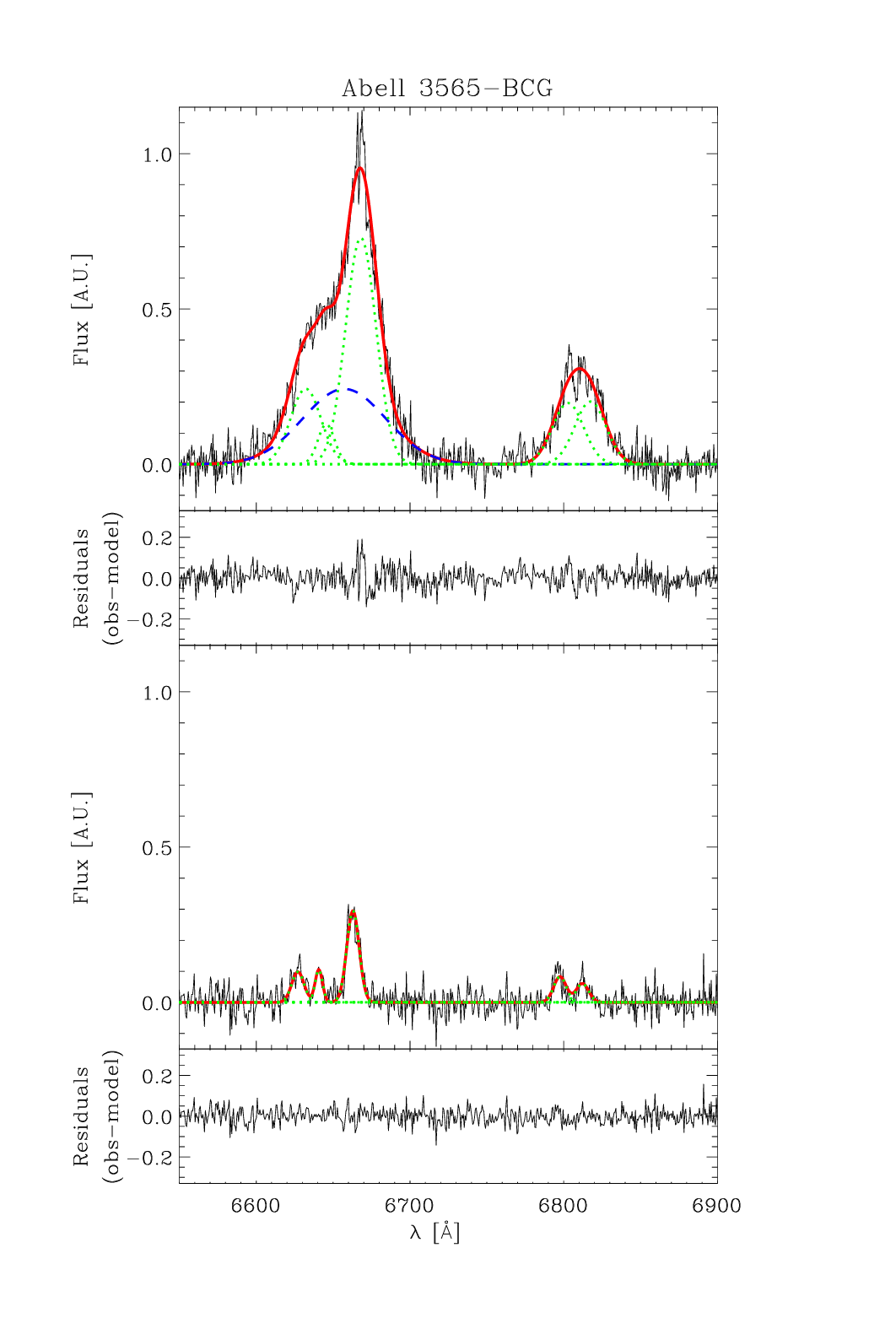}
\caption{{\small As in Figure~\ref{1836spec}, but for Abell~3565-BCG.
  The outer spectrum was extracted at $r=+0\Sec25$.}
\label{3565spec}}
\end{figure}

\clearpage

\begin{figure}
\epsscale{.7}
\plotone{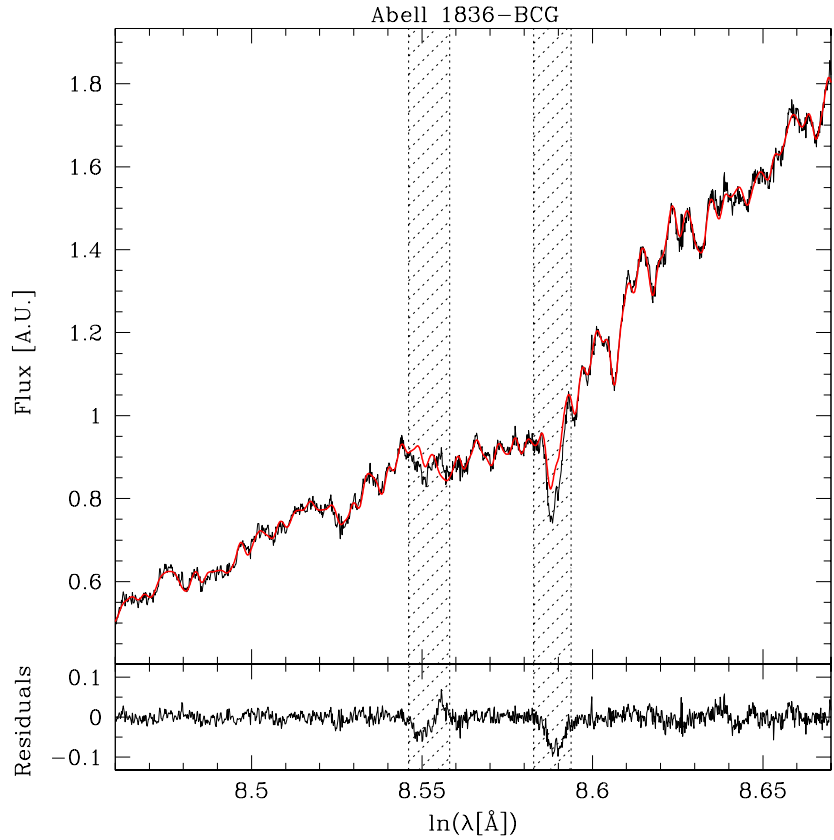}
\caption{ Best fit of the broadened spectrum of the template star
  (thick red line) to the spectrum of Abell~1836-BCG (thin black line) 
  used to measure the galaxy velocity dispersion. The
  hatched regions mark the spectral ranges excluded from the fit to
  minimize the mismatch of the abundance ratios. The difference
  between the galaxy spectrum and broadedned template spectrum is
  shown in the bottom panel.}
\label{fig:sigma}
\end{figure}

\clearpage

\begin{figure}
\epsscale{.35}
\plotone{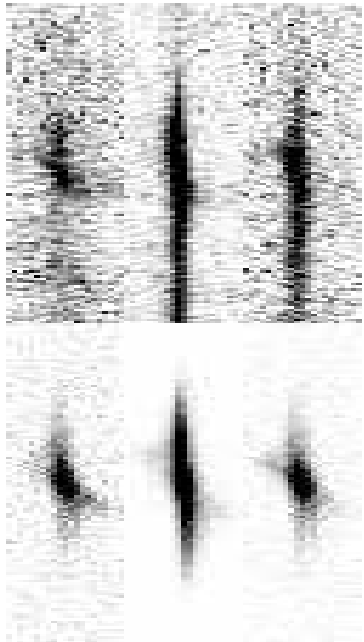}
\caption{{\small Comparison between the bidimensional observed 
    ({\em upper panels}) and synthetic spectra ({\em
    lower panels}) for the best-fitting model of Abell
    1836-BCG. Spectra obtained in position 1 ({\em central panels}), 2
    ({\em right panels}), and 3 ({\em left panels}) are shown. In the
    models we added a random noise to mimic the actual S/N of
    the observed spectra.}
\label{mod2d1836}}
\end{figure}

\begin{figure}
\epsscale{1.}
\plotone{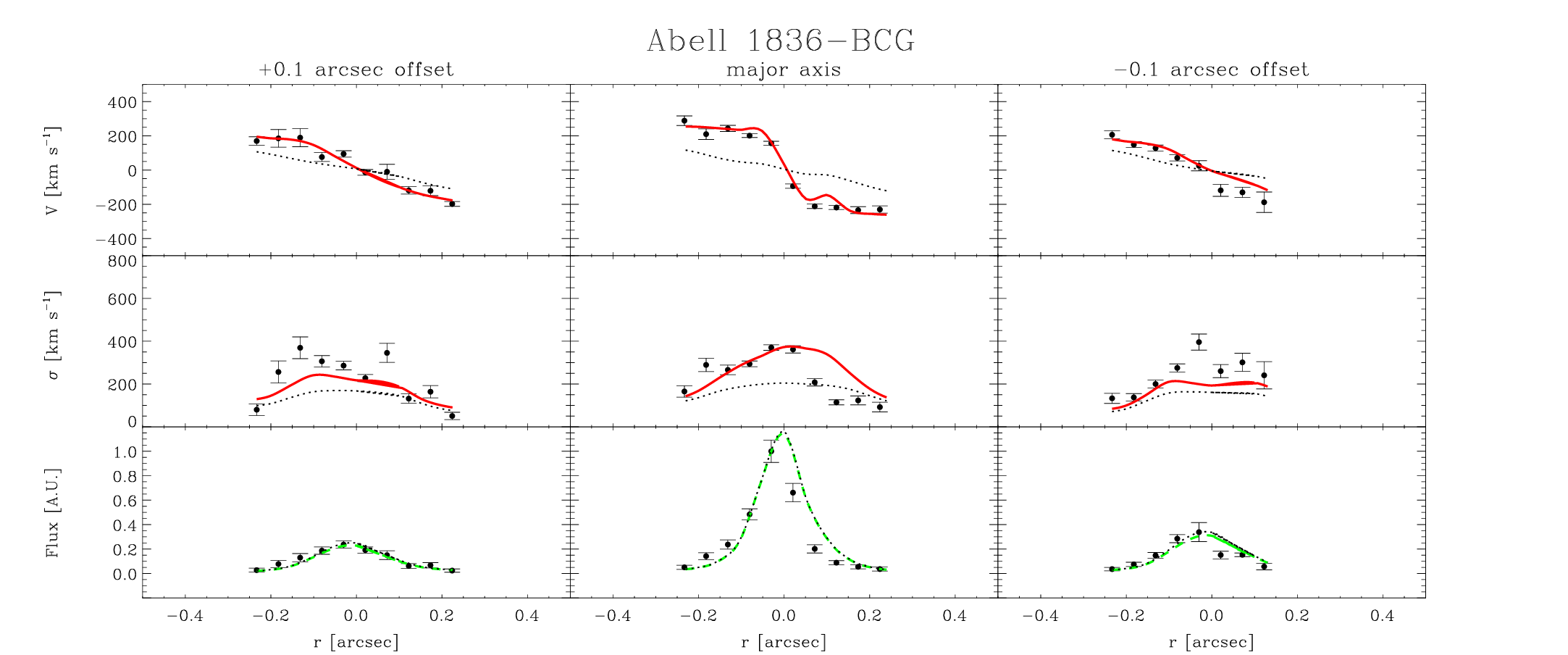}
\caption{{\small Observed \niig\ kinematics ({\it filled circles\/})
    along with the best-fitting model ({\it solid lines\/}) for the
    SBH mass of Abell~1836-BCG. The observed and modeled velocity
   curve ({\em top panels}), velocity dispersion radial profile
   ({\em central panels}), and flux profile ({\em bottom panels}) are
   shown for the slit along the $+0\Sec1$ offset position ({\em
     left panels}), the major axis ({\em central panels}), and the
   $-0\Sec1$ offset position ({\em right panels}) of the gas disk.
   The dotted lines correspond to a model obtained with $M_\bullet=0$ \msun,
$i=69.0\dg$, and \mlstar $=5.0$ \mlsun (in the $I-$band).}}
\label{obs_mod1836}
\end{figure}

\begin{figure}
\epsscale{.75}
\plotone{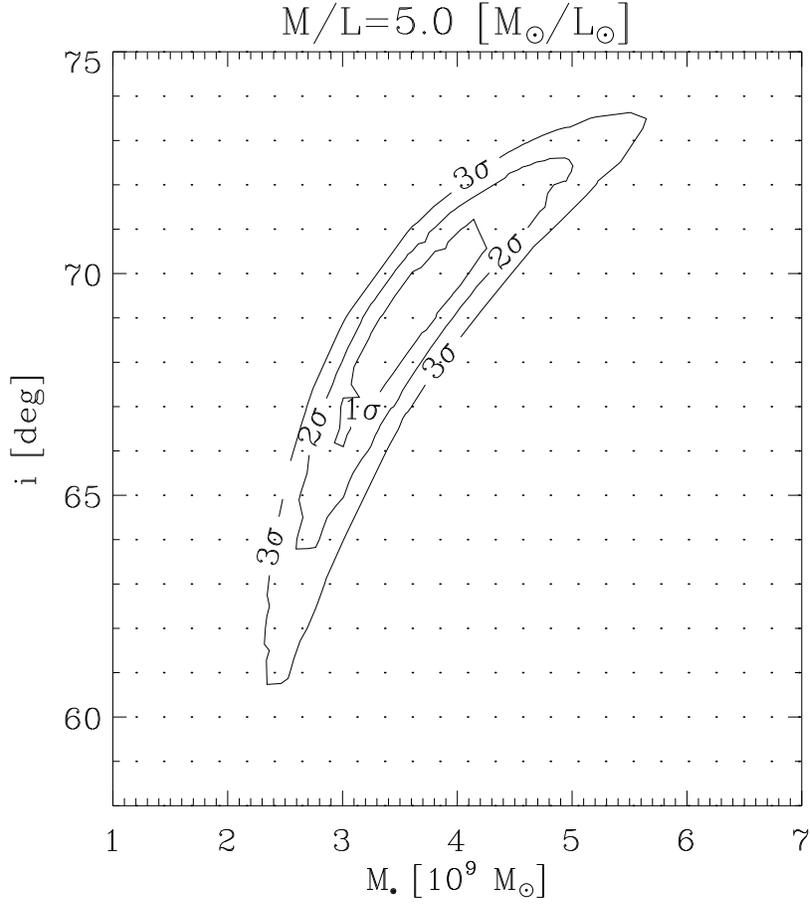}
\caption{{\small Locus of points of equal $\chi^2$ values around the minimum $\chi^2_m$ value for the model of Abell~1836-BCG. The $1\sigma$, $2\sigma$,
and $3\sigma$ confidence levels expected for two free parameters are shown in the \mbh-$i$ plane,
with \mlstar$=5.0$ \mlsun\, corresponding to the $\chi^2_m$ value. The grid of models explored is shown by the dots.}} 
\label{contour1_1836}
\end{figure}

\begin{figure}
\epsscale{.75}
\plotone{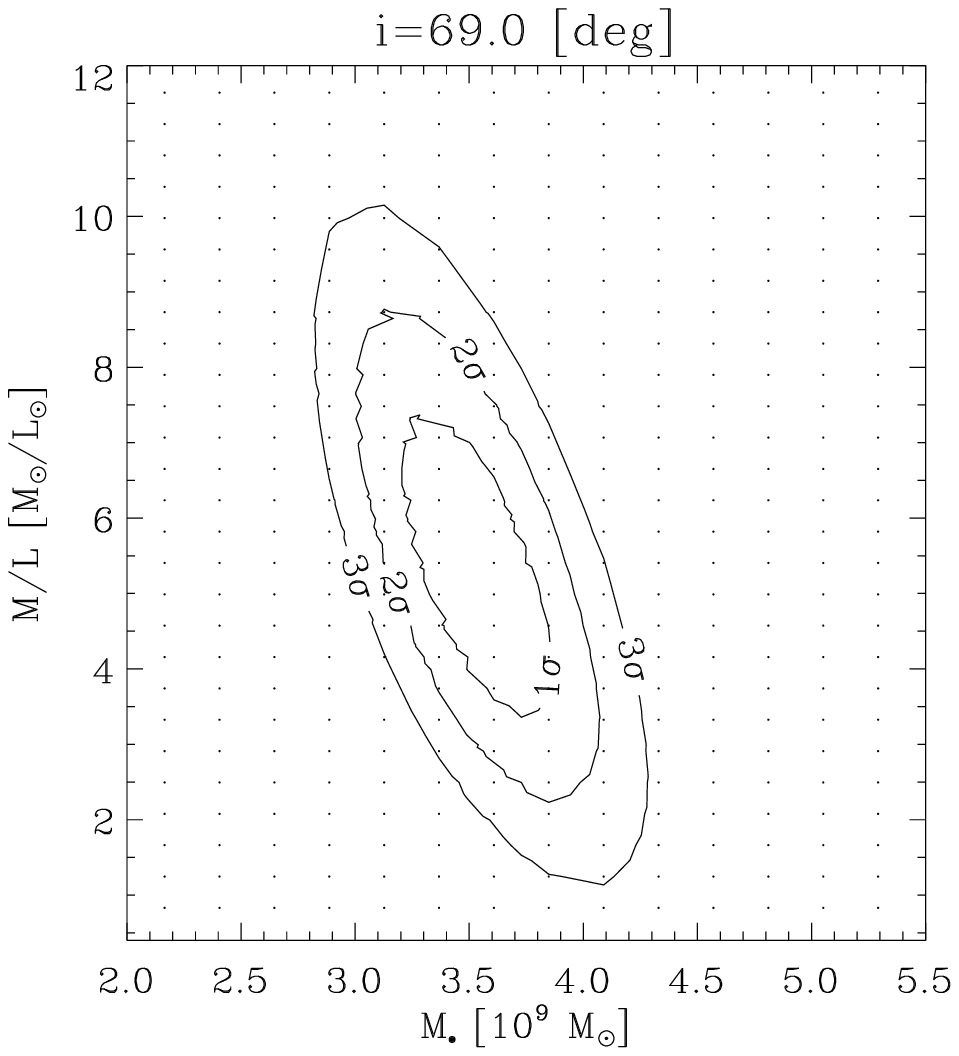}
\caption{{\small As in Figure~\ref{contour1_1836}, but for the $1\sigma$, $2\sigma$,
and $3\sigma$ confidence levels expected for two free parameters shown in the \mbh-\mlstar\ plane with $i=69^\circ$ corresponding to the $\chi^2_m$ value.}} 
\label{contour2_1836}
\end{figure}

\begin{figure}
\epsscale{.75}
\plotone{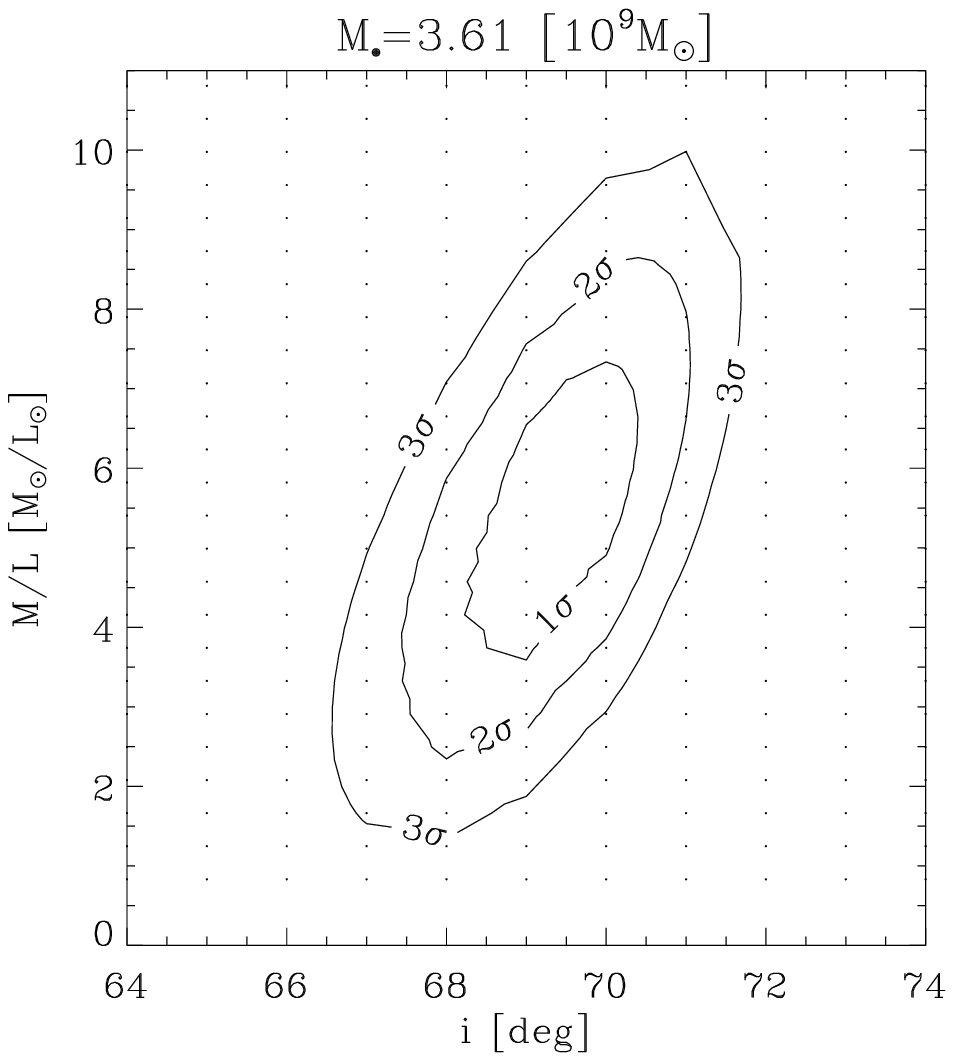}
\caption{{\small As in Figure~\ref{contour1_1836}, but for the  $1\sigma$, $2\sigma$,
and $3\sigma$ confidence levels expected for two free parameters shown
in the $i$-\mlstar\ plane with \mbh$=3.61\cdot 10^9$ \msun corresponding to the $\chi^2_m$ value.}} 
\label{contour3_1836}
\end{figure}

\begin{figure}
\epsscale{.8}
\plotone{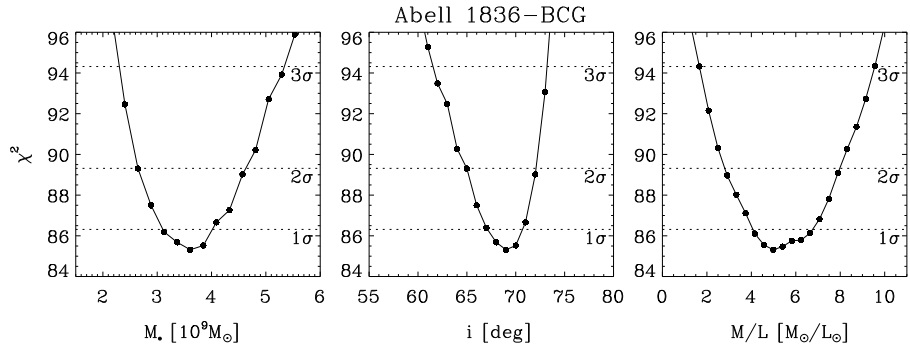}
\caption{{\small $\chi^2$ distribution for Abell~1836-BCG as a
    function of \mbh\ ({\em left}), inclination ({\em center}), and \mlstar\ ({\em right}). The dotted horizontal
    lines indicate the confidence levels on the best fitting values,
    marginalizing over all other parameters.}} 
\label{sigma_est1836}
\end{figure}

\clearpage

\begin{figure}
\epsscale{.45}
\plotone{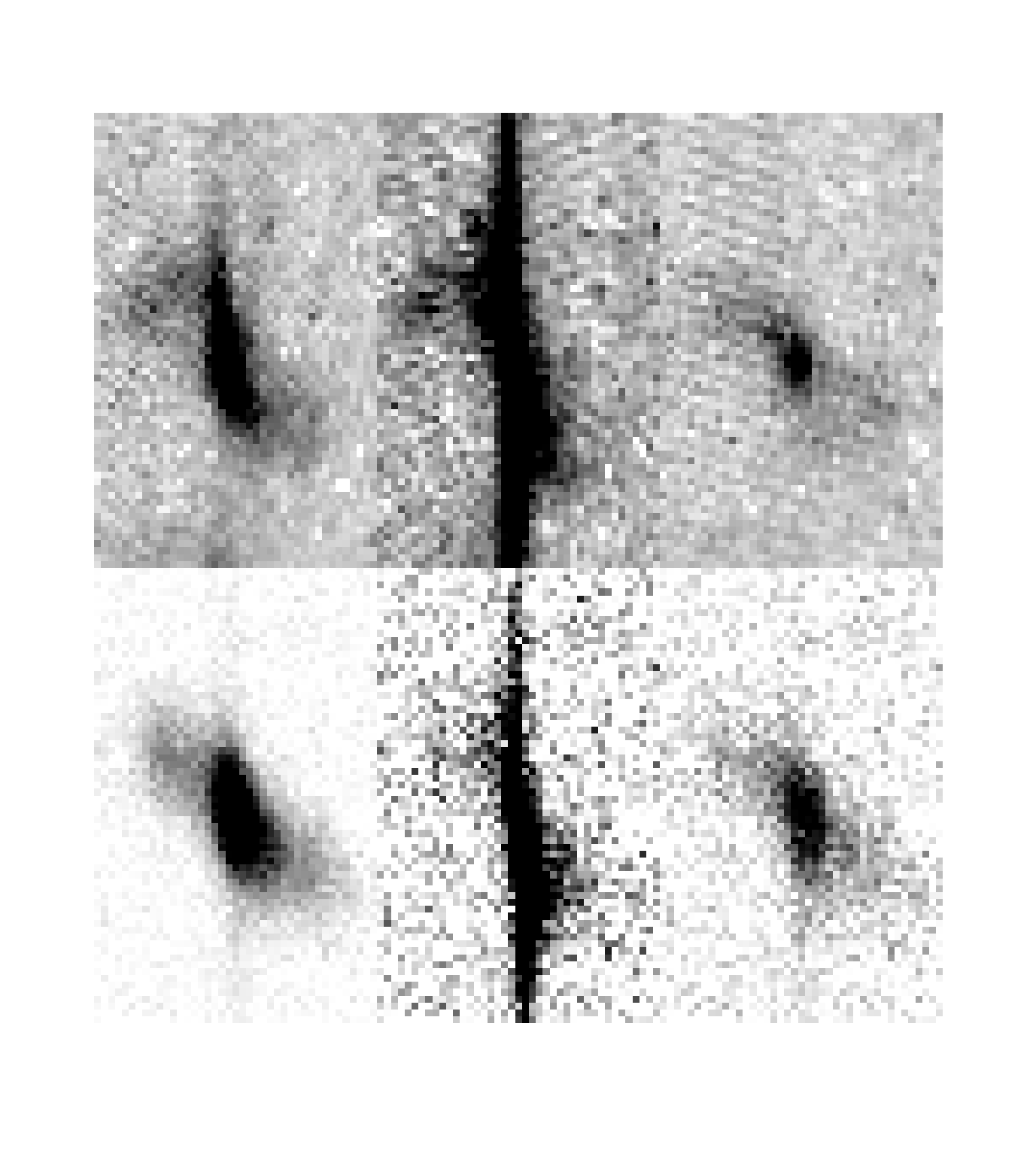}
\caption{{\small As in Figure~\ref{mod2d1836}, but for Abell~3565-BCG.}
\label{mod2d3565}}
\end{figure}

\begin{figure}
\epsscale{1.}
\plotone{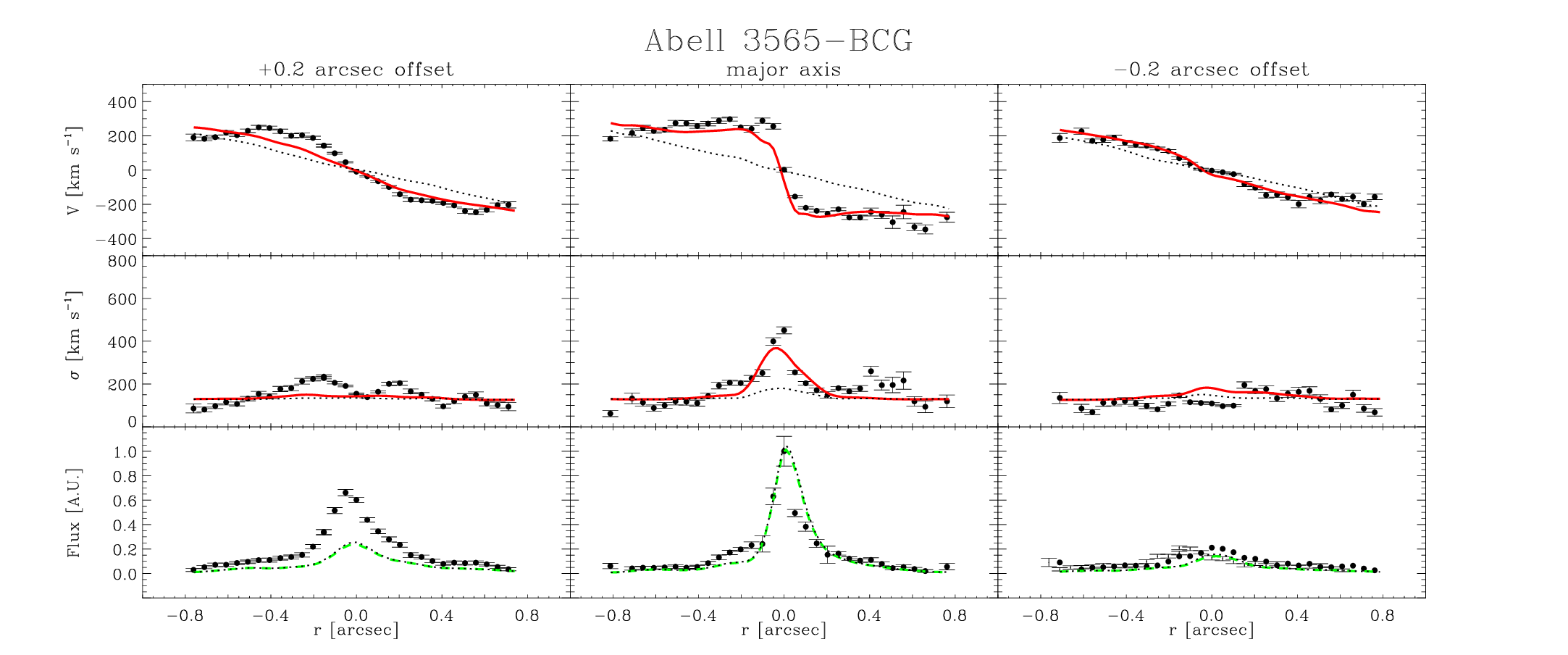}
\caption{{\small Observed \niig\ kinematics ({\it filled circles\/})
along with the best-fitting model ({\it solid lines\/}) for the SBH
mass of Abell~3565-BCG.   The observed and modeled velocity
   curve ({\em top panels}), velocity dispersion radial profile
   ({\em central panels}), and flux profile ({\em bottom panels}) are
   shown for the slit along the $+0\Sec1$ offset position ({\em
     left panels}), the major axis ({\em central panels}), and the
   $-0\Sec1$ offset position ({\em right panels}) of the gas disk.
   The dotted lines correspond to a model obtained with $M_\bullet=0$ \msun,
$i=66.0\dg$, and \mlstar $=6.3$ \mlsun (in the $I-$band).}}
\label{obs_mod3565}
\end{figure}

\clearpage

\begin{figure}
\epsscale{.75}
\plotone{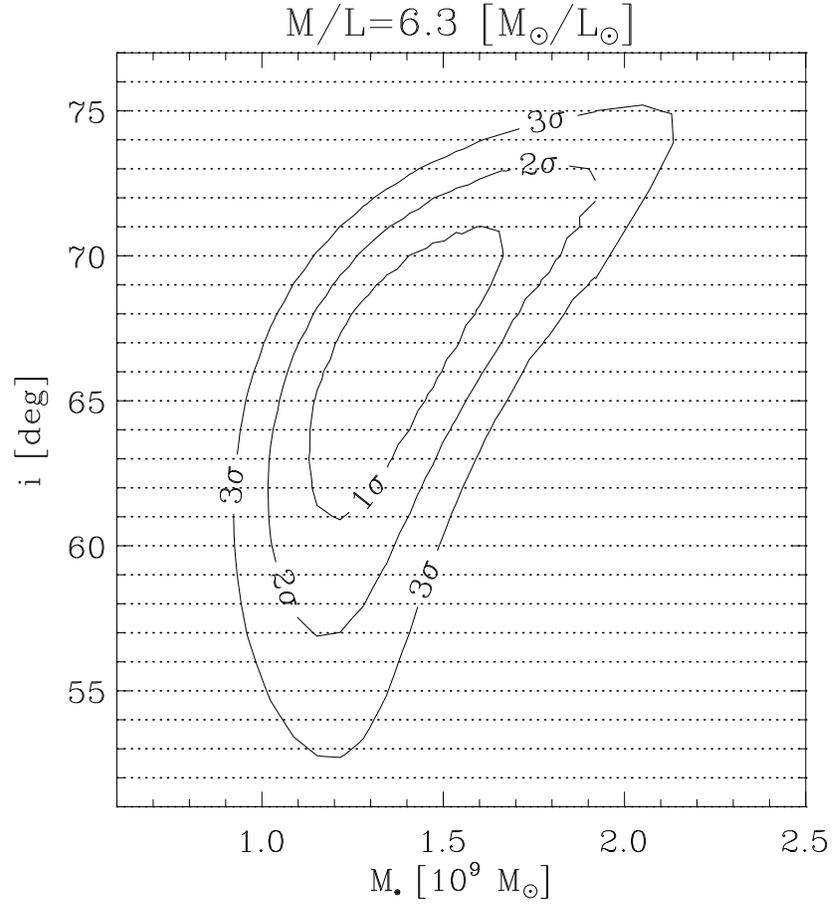}
\caption{{\small Locus of points of equal $\chi^2$ values around the minimum $\chi^2_m$ value for the modeling of Abell~3565-BCG. The $1\sigma$, $2\sigma$,
and $3\sigma$ confidence levels expected for two free parameters are shown in the \mbh-$i$ plane,
with \mlstar$=6.3$ \mlsun\, corresponding to the $\chi^2_m$ value.}} 
\label{contour1_3565}
\end{figure}

\begin{figure}
\epsscale{.75}
\plotone{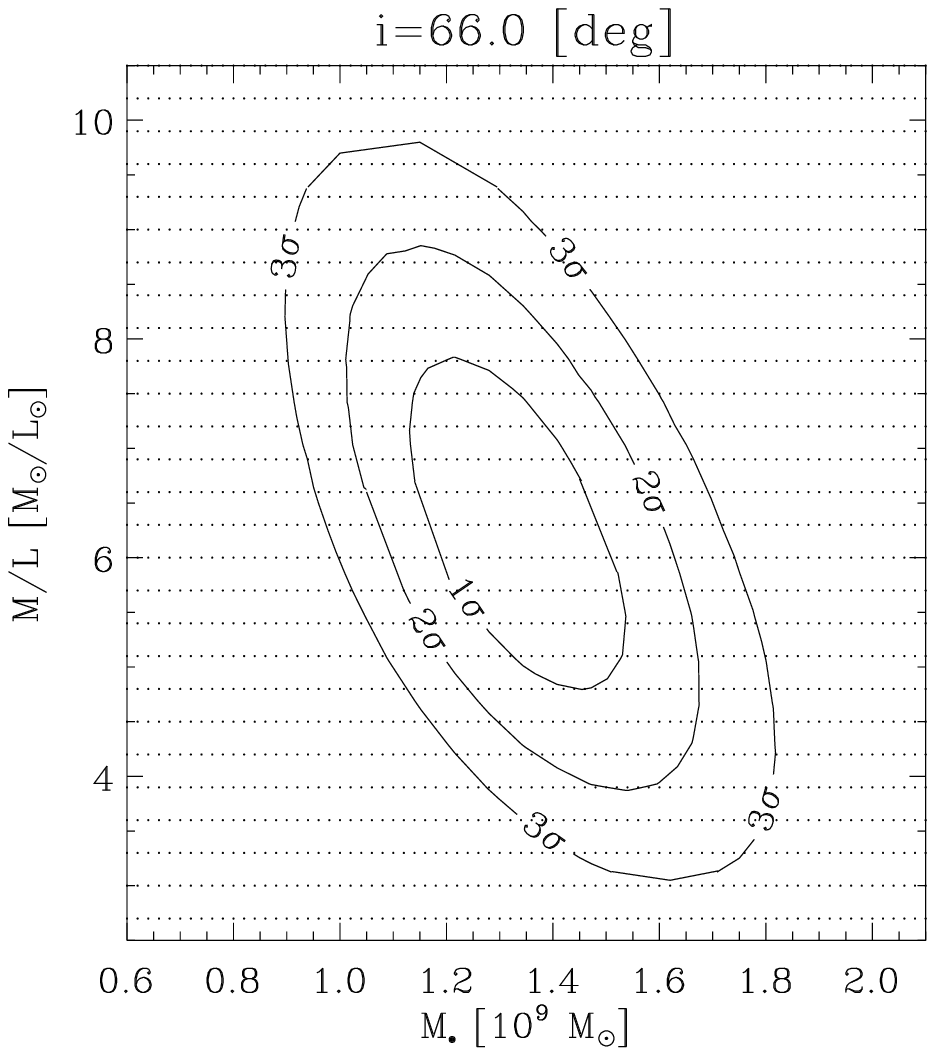}
\caption{{\small As in Figure~\ref{contour1_3565}, but for the $1\sigma$, $2\sigma$,
and $3\sigma$ confidence levels expected for two free parameters shown in the \mbh-\mlstar\ plane with $i=66^\circ$ corresponding to the $\chi^2_m$ value.}} 
\label{contour2_3565}
\end{figure}

\begin{figure}
\epsscale{.75}
\plotone{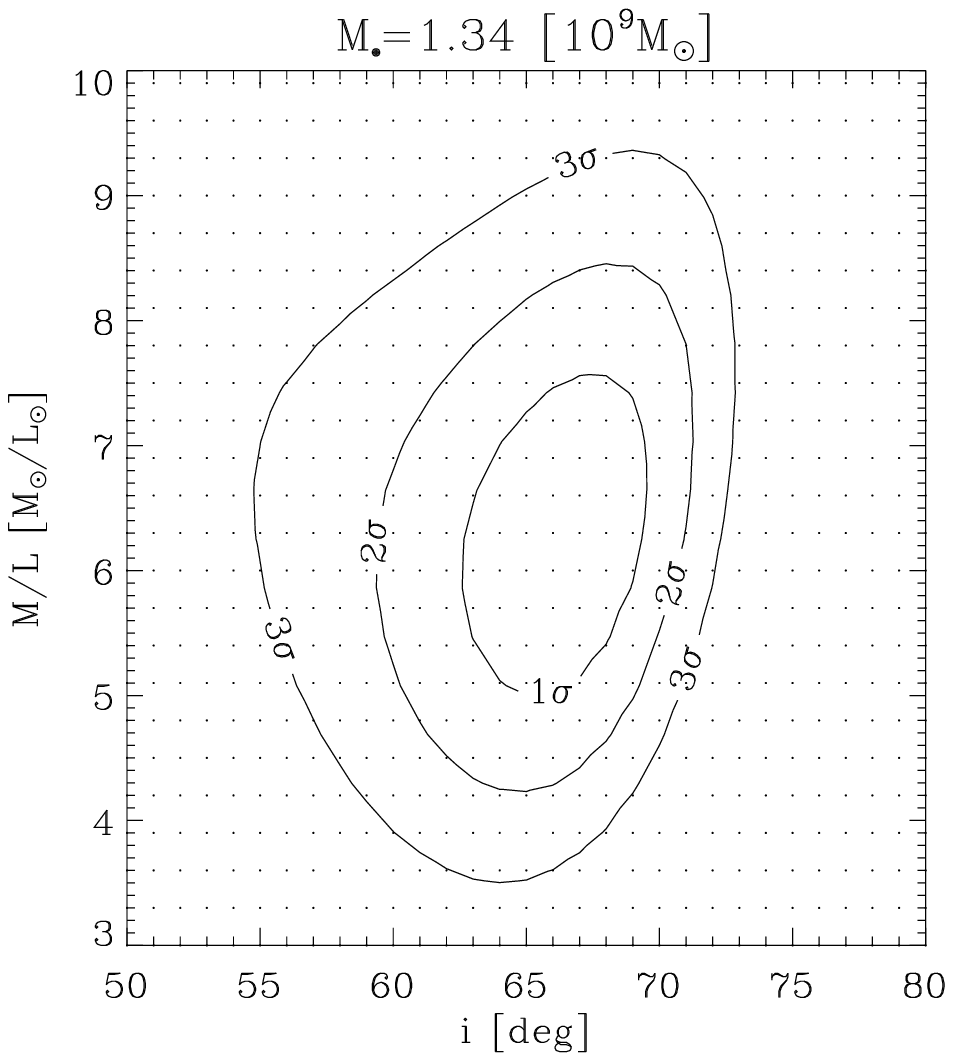}
\caption{{\small As in Figure~\ref{contour1_3565}, but for the $1\sigma$, $2\sigma$,
and $3\sigma$ confidence levels expected for two free parameters shown
in the $i$-\mlstar\ plane with \mbh$=1.34\cdot 10^9$ \msun corresponding to the $\chi^2_m$ value.}} 
\label{contour3_3565}
\end{figure}

\clearpage

\begin{figure}
\epsscale{.8}
\plotone{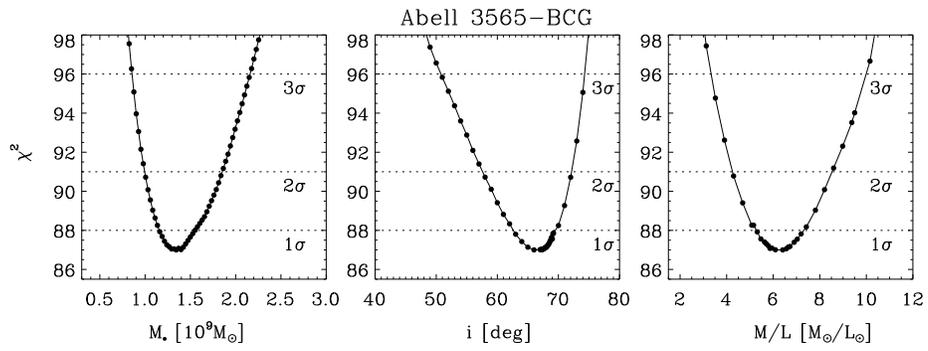}
\caption{{\small Same of Figure~\ref{sigma_est1836}, but for Abell~3565-BCG.}} 
\label{sigma_est3565}
\end{figure}

\clearpage

\begin{figure}
\epsscale{.7}
\plotone{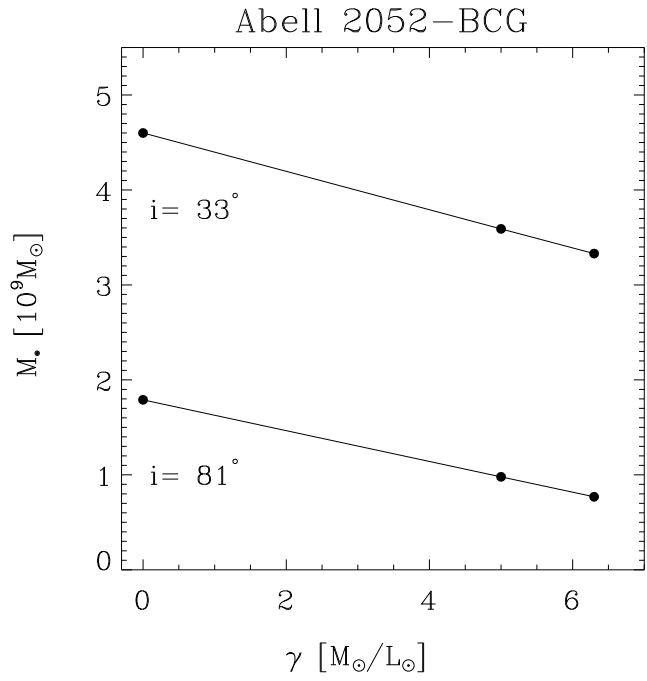}
\caption{Upper limit of SBH mass of Abell~2052-BCG as function of the
  \mlstar\ assumed, for an inclination of the gas-disk $i= 33\dg$ and $i=81\dg$.} 
\label{upper_limit}
\end{figure}

\clearpage

\begin{figure}
\epsscale{1.}
\plotone{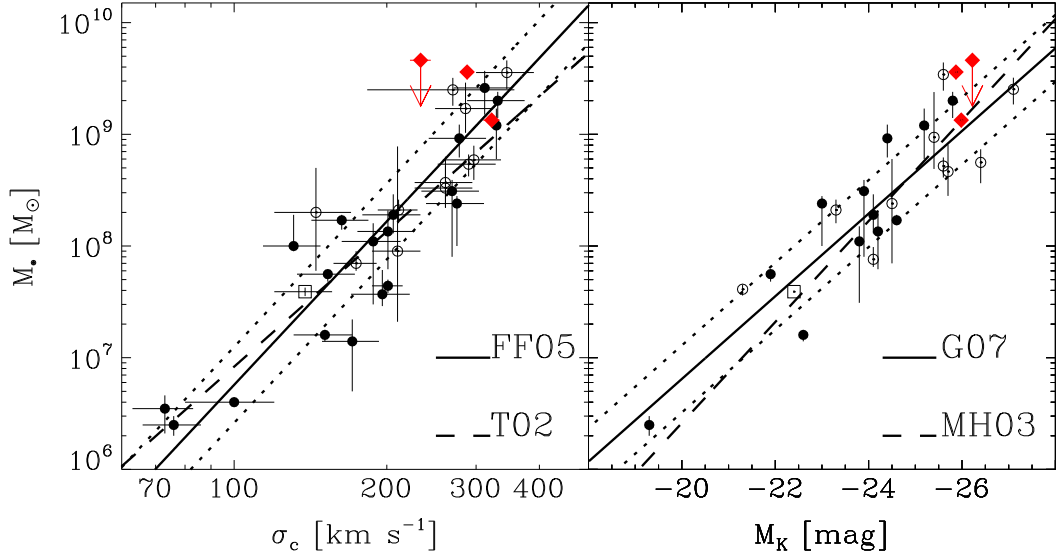}
\caption{\small {\it Left panel\/}: Location of the SBHs masses of our
BCG sample galaxies ({\em diamonds}) with respect to the
\mbh$-$\sigmac\ relation by Ferrarese \& Ford (2005).  We plot the SBH
masses based on resolved dynamical studies of ionized gas ({\it open
circles\/}), water masers ({\it open squares\/}), and stars ({\it
filled circles\/}) from Table 2 of Ferrarese \& Ford (2005). The
dashed line represents the Tremaine et al. (2002) relation.  {\it
Right panel\/}: Location of our BCG sample galaxies with respect to
the near-infrared ($K$-band) \mbh$-$\lbulge\ relation by Graham
(2007).  Data are taken from Table 2 of Graham (2007) and symbols are
as in the left panel. Masses and magnitudes of NGC 5252, NGC 6251 and
Cygnus A were adjusted to a distance obtained with $H_0=75$ km
s$^{-1}$ Mpc$^{-1}$.  The dashed line represents the Marconi \& Hunt
(2003) relation.  In both panels, we added the data relative to NGC
1399 (Houghton et al. 2006) and the dotted lines represent the
$1\sigma$ scatter in \mbh.\label{scal_rel}}
\end{figure}
\clearpage

\end{document}